\documentclass[graybox, nosecnum]{svmult}

\usepackage{mathptmx}       % selects Times Roman as basic font
\usepackage{helvet}         % selects Helvetica as sans-serif font
\usepackage{courier}        % selects Courier as typewriter font
\usepackage{type1cm}        % activate if the above 3 fonts are
\usepackage{makeidx}         % allows index generation
\usepackage{graphicx}        % standard LaTeX graphics tool
\usepackage{multicol}        % used for the two-column index
\usepackage[bottom]{footmisc}% places footnotes at page bottom
\usepackage{hyperref}        %for hyperlinks
\usepackage{soul}            % for high-lighting of text

\newcommand{\be}{\begin{equation}}
\newcommand{\ee}{\end{equation}}
\newcommand{\bea}{\begin{eqnarray}}
\newcommand{\eea}{\end{eqnarray}}

\newcommand{\lambdabar}{{\mkern0.75mu\mathchar '26\mkern -9.75mu\lambda}}
\makeindex             % used for the subject index
\begin{document}
%\tableofcontents{}
\title*{Theoretical description of pygmy (dipole) resonances}
\author{Edoardo G. Lanza \thanks{corresponding author} and Andrea Vitturi}
\institute{Edoardo G. Lanza \at INFN, Sezione di Catania, via Santa Sofia 64, I-95123 Catania, Italy, \\\email{edoardo.lanza@ct.infn.it}
\and Andrea Vitturi \at Dipartimento di Fisica e Astronomia ``Galileo Galilei", Universit\`a di Padova, and INFN, Sezione di Padova, via Francesco Marzolo 8, 35121 Padova, Italy, \\
\email{andrea.vitturi@pd.infn.it}}
\maketitle
\abstract{ Stable and unstable nuclei with neutron excess ($N>Z$) show - in the isovector dipole transition strength distribution - a small hump around the neutron emission threshold energy known as Pygmy Dipole Resonance (PDR). One of its main features is the isospin mixing allowing the experimental studies with both isovector and isoscalar probes. Different theoretical approaches and methodologies are used to deduce the characteristics of the PDR. In this Chapter, the various mean-field theories and their extensions, devoted to understand and reproduce the strength distribution of these low-lying dipole states, are summarised. Special attention is dedicated to the calculations of the inelastic cross section, aspect that is particularly important in the investigation with isoscalar probes, such as $\alpha$-particles or $^{17}$O. The relevance of the radial form factors is presented in relation to the inelastic cross-section calculations. 
}

\section{\textit{Introduction}}

States that can be interpreted as the quanta of collective vibrations are a general property of quantum mesoscopic systems, which can be found in various fields of physics. In nuclear physics, such vibrational states of the nucleus  have been known for many years~\cite{Boh75}. Among them, the Giant Resonances (GRs) have attracted most of the attention because they appear as broad resonances and are ubiquitous along the whole Segre's diagram \cite{Bor19,Har01}. Giant resonances are the result of a collective motion of many nucleons in the nucleus. The description of the GR is commonly done in terms of quasi-harmonic vibration around the ground-state density of the nucleus or through quantum-mechanical transition from the ground state to the collective one. In both cases, the amount of strength used by the mode is the predominant part of the sum rule whose value is determined by the ground-state properties of the nucleus. The Isovector Giant Dipole Resonance (IVGDR) was the first one discovered and the most studied one. The macroscopic model of Goldhaber and Teller (GT) \cite{Gol48} and of Steinwedel and  Jensen (SJ) \cite{Ste50} describe the mode as a collective out-of-phase motion of all the protons against all the neutrons. Responsible for this excitation is the electromagnetic interaction capable also to excite different types of vibrational modes that can be classified in terms of spin, isospin and multipolarity \cite{Bor19,Har01}. These vibrations are called isoscalar or isovector depending whether the oscillations of neutrons and protons are in phase or out of phase, respectively. The centroid, the strength and the width of the Giant Resonances - the most relevant properties of the GR - depend on the bulk structure of the nuclei. For the isovector GDR (IVGDR), the energy centroid of the strength distribution changes, as function of the mass number $A$, according to the expression $E_{x}= 31.2 \> A^{-1/3} + 20.6\> A^{-1/6}$.

Nuclei with the number of neutrons (N) greater than the number of protons (Z) show - in the isovector dipole transition strength distribution - an additional small hump around the neutron emission threshold energy. This is valid for stable and unstable nuclei and since the first observation, this structure was referred to as Pygmy Dipole Resonances (PDR). The name comes from the small percentage of the energy-weighted sum rule (EWSR) compared to the one exhausted by the IVGDR. In the last two decades, numerous experimental and theoretical work has been dedicated to investigate this excitation mode - interesting for itself - with also some important spin-offs on some other physics fields. A complete overview of the problem can be obtained by reading the recently published reviews \cite{Paa07,Sav13,Bra15,Bra19,Aum19}. Few and precise features can be extracted from these investigations: they are present only in nuclei with neutron excess, they have been measured below and above the neutron emission threshold and they have a strong isospin mixing. This latter characteristic has allowed their studies using isoscalar probes, such as $\alpha$ particles via the nuclear interaction force. These investigation methods, combined with the classical electromagnetic ($\gamma, \gamma^{\prime}$) reaction, revealed unexpected behaviour of the dipole states in the energy region below the neutron emission threshold: the low-lying states are excited by both isoscalar and isovector probes while the ones in the higher energy region are populated only by the electromagnetic field. This phenomenon is now identified as isospin (or PDR) splitting (see Ref. \cite{Sav13} and references therein). It is present in all the nuclei with $N>Z$ that have been investigated and it has been confirmed by measurements done with other isoscalar probes as $^{17}$O  or ($p,p^{\prime}$) at tens of MeV reactions (see Ref. \cite{Bra15,Bra19} and references therein). It is not known whether this isospin splitting is also present at the energy region above the neutron emission threshold. An early attempt was done in Ref. \cite{Mar18} where the unstable isotope $^{68}$Ni was used as projectile on a $^{12}$C target. The isospin splitting of the PDR is not evident from the available data. Therefore, more experiments with better energy resolution and statistics are needed in order to clarify this point. Other aspects need further attention such as the collective nature of this mode or the understanding of the interplay between the isoscalar and isovector contribution. 

Many theoretical works employing different approaches and methodologies are used to deduce the characteristics of the PDR. The GT and SJ macroscopic models have been extended to take into account explicitly the existence of the neutron excess that often has been considered as a kind of skin surrounding an isospin inert core. Microscopic approaches use all the variation of mean-field theories from the Random-Phase Approximation (RPA) to the quasiparticle RPA (QRPA) with its relativistic version (RQRPA) up to the theories that take into account the coupling of particle-hole excitations with more complicated configurations like two- or three-phonon states. The aim of these theoretical approaches is to obtain a good description of the measured dipole strength distribution. Many of these calculations have succeeded in the reproduction and understanding of the experimental data. When the combined isovector and isoscalar probes are used then it is important also  to calculate the cross section being the quantity measured when the excitation is due to the isoscalar nuclear interaction. 

An  overview of the main theoretical approaches will be given in this Chapter with particular attention to the calculation of the inelastic cross section. 

\section{\textit{Experimental evidences}\label{exp}} 

The most natural way to extract information for the low-lying dipole states is to look at the response to the electromagnetic interaction in the ($\gamma, \gamma^{\prime}$) reactions. This excitation mechanism is often identified as Nuclear Resonance Fluorescence (NRF) or photon scattering where a photon beam is absorbed by the target nucleus to an excited state, which decays by $\gamma$ emission. The photon beams can be generated by bremsstrahlung produced by the interaction between an electron beam with a radiator material. Another method to obtain a quasi-mono-energetic photon beam is to use the Compton scattering of laser photons off ultra-relativistic electrons in a storage ring. A beam of protons accelerated to relativistic energy and inelastically scattered and detected at around $0^{\circ}$ scattering angle can also be used due to the high selectivity of the Coulomb contribution in the total excitation cross section.

The nature of the PDR can be investigated via the inelastic scattering of light-nuclei at intermediate energy. These reactions explore the surface nuclear region while the photon beam interacts with the entire bulk of the nucleus. The inelastically scattered particle and the $\gamma$ emitted by the PDR are measured in coincidence and their angular correlation allows the identification of the multipolarity of the $\gamma$ rays. For unstable neutron-rich nuclei, inverse kinematic experiments have to be used. The incoming beam is subjected to in-flight fragmentation yielding a secondary beam, which is extracted using fragment separators. Detecting the residual nuclei, neutrons and $\gamma$ rays following the scattering of the secondary beam the initial excitation energy is reconstructed by the invariant-mass method. It is also possible to measure directly the $\gamma$ decay from the excited nucleus imposing that the residual fragment measured in coincidence is the same as the projectile. 

Other more sophisticated methods are employed to study the PDR like the neutron resonance scattering or low-energy light-nuclei scattering to extract the gamma strength function or via detecting the $\gamma$ rays following $\beta$ decays to deduce the decay branching ratio. For more details on the different experimental methods and the related results, it is suggested to consult the chapters of this Handbook by Zilges and Savran or the reviews \cite{Sav13,Bra19} and references therein.

The experimental evidences of the PDR have been found above and below the neutron emission threshold. The presence of the PDR above the neutron separation energy has been measured with the virtual-photon excitation method at GSI for the exotic neutron-rich nuclei $^{130,132}$Sn \cite{Adr05, Kli07} and $^{68}$Ni \cite{Wie09,Ros13} with relativistic Coulomb excitation. Other exotic nuclei, like $^{20,22}$O ,$^{26}$Ne, $^{70}$Ni, have shown a similar structure in the low-energy tail of the IVGDR strength distribution. Their percentage of the EWSR is a few percent and their summed strength increases with the $N/Z$ ratio. More details can be found in the reviews \cite{Sav13,Bra19} and reference therein. 
The results of a first measurement of a relativistic Coulomb excitation of the two unstable Sn isotopes mentioned above \cite{Adr05} are shown in Fig. 9 of the Chapter by Zilges and Savran on the experimental studies of the PDR. The mass-invariant method was used to deduce the excitation energy of the projectile. The photo-neutron cross section was extracted, from the measured electromagnetic cross section, using the virtual-photon method. The nuclear contribution was subtracted as well as the one coming from the ISGQR. The peaks of the PDR are visible at energies just above the neutron emission threshold.

\begin{figure}[!htb]
\begin{center}
\includegraphics[width=8cm, angle=0]{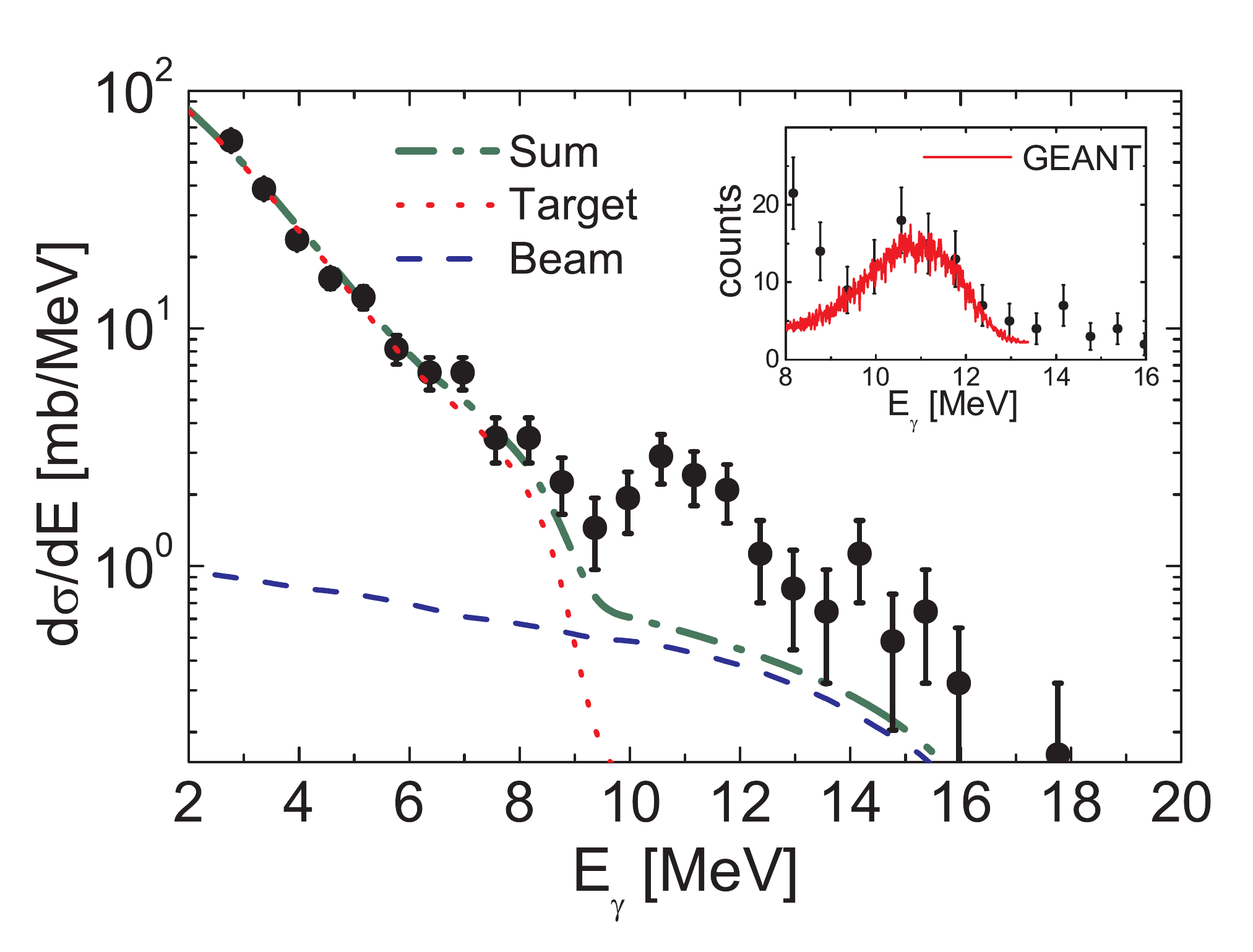}
\caption{(Color online) The $\gamma$-ray spectrum for the reaction $^{68}$Ni on a Au target at 600 MeV/u. The dashed  and dotted lines are statistical-model calculations for the projectile and target, respectively.  In the inset, the GEANT \cite{GEANT4} simulation for a transition at 11 MeV is shown as solid red line. Taken with permission from \cite{Wie09}. \copyright 2009 by APS.}
\label{fig-wie09}
\end{center}
\end{figure}
The relativistic Coulomb excitation of $^{68}$Ni in the reaction with a Au target also shows an enhancement slightly above the neutron emission threshold as it is shown in Fig. \ref{fig-wie09} where the $\gamma$-ray spectrum is plotted together with statistical-model predictions for the target (dotted line) and the beam (dashed line). The peak at 11 MeV correspond to the pygmy dipole resonance as validated by the GEANT simulation calculation shown in the inset of Fig. \ref{fig-wie09}.

\begin{figure}[!htb]
\begin{center}
\includegraphics[width=8cm, angle=0]{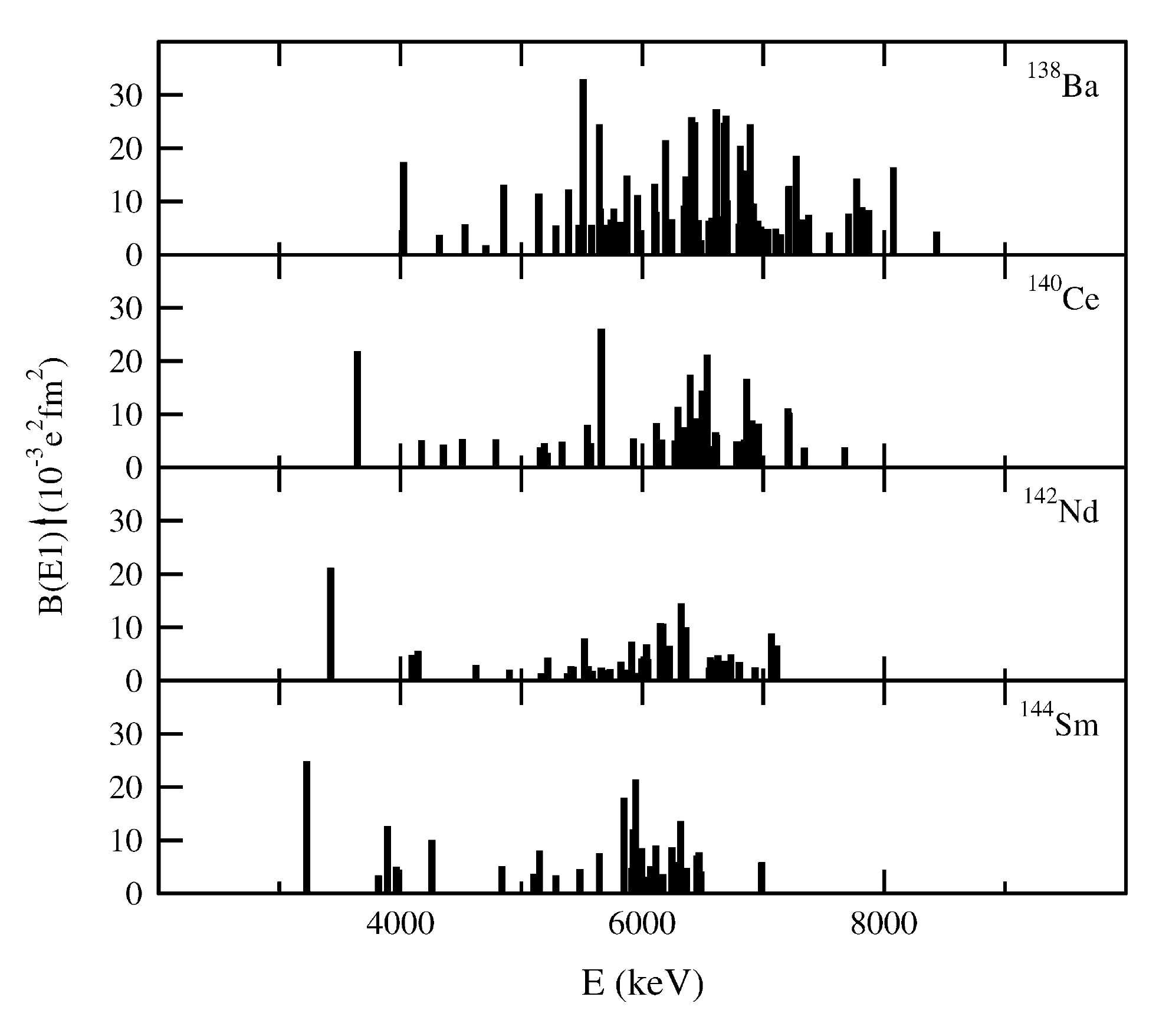}
\caption{Dipole reduced transition probabilities B(E1) measured in a photon scattering experiment for five stable even $N$=82 isotones. Taken with permission from \cite{Vol06}. \copyright 2006 by Elsevier.}
\label{fig-vol06}
\end{center}
\end{figure}
Most of the experimental investigations for the PDR have been done for stable nuclei with neutron excess. A  concentration of E1 strength was observed below the neutron-separation threshold in $N$=82 stable isotones  \cite{Zil02,Vol06,Sav11} and around the neutron threshold in $^{208}$Pb \cite{End00,Rye02,End03}. As an example, the dipole reduced transition probabilities B(E1) for five stable even $N$=82 isotones are shown in Fig. \ref{fig-vol06}. The NRF technique allows to resolve all the single dipole states below the neutron emission threshold. The dipole nature of the gamma rays is established by the measure of the ratio of the $\gamma$-rays intensity at two different angles \cite{Vol06}. 

One of the main features of the Pygmy Dipole Resonances is the isospin mixing - as will be shown in some details in the next section - that manifests in  the shape of the transition density for this excitation mode. This property  allows the possibility to excite the PDR also via an isoscalar probe, that is with a nuclear interaction. Indeed, the excitation of the low-lying dipole states using light ions in reaction as ($p,p^{\prime} \gamma$), ($\alpha,\alpha^{\prime} \gamma$) and  ($^{17}$O,$^{17}$O$^{\prime} \gamma$) have been employed.  The particle(recoil)-$\gamma$ coincidence allows the selection of the E1 strength via the angular correlation $W(\theta_\gamma)$ of the outgoing products. The comparison between these measurements and the results from the ($\gamma,\gamma^{\prime}$) reactions shows an unexpected feature of the low-lying dipole strength distribution below the neutron emission threshold. 
\begin{figure}[!htb]
\begin{center}
\includegraphics[width=8cm, angle=0]{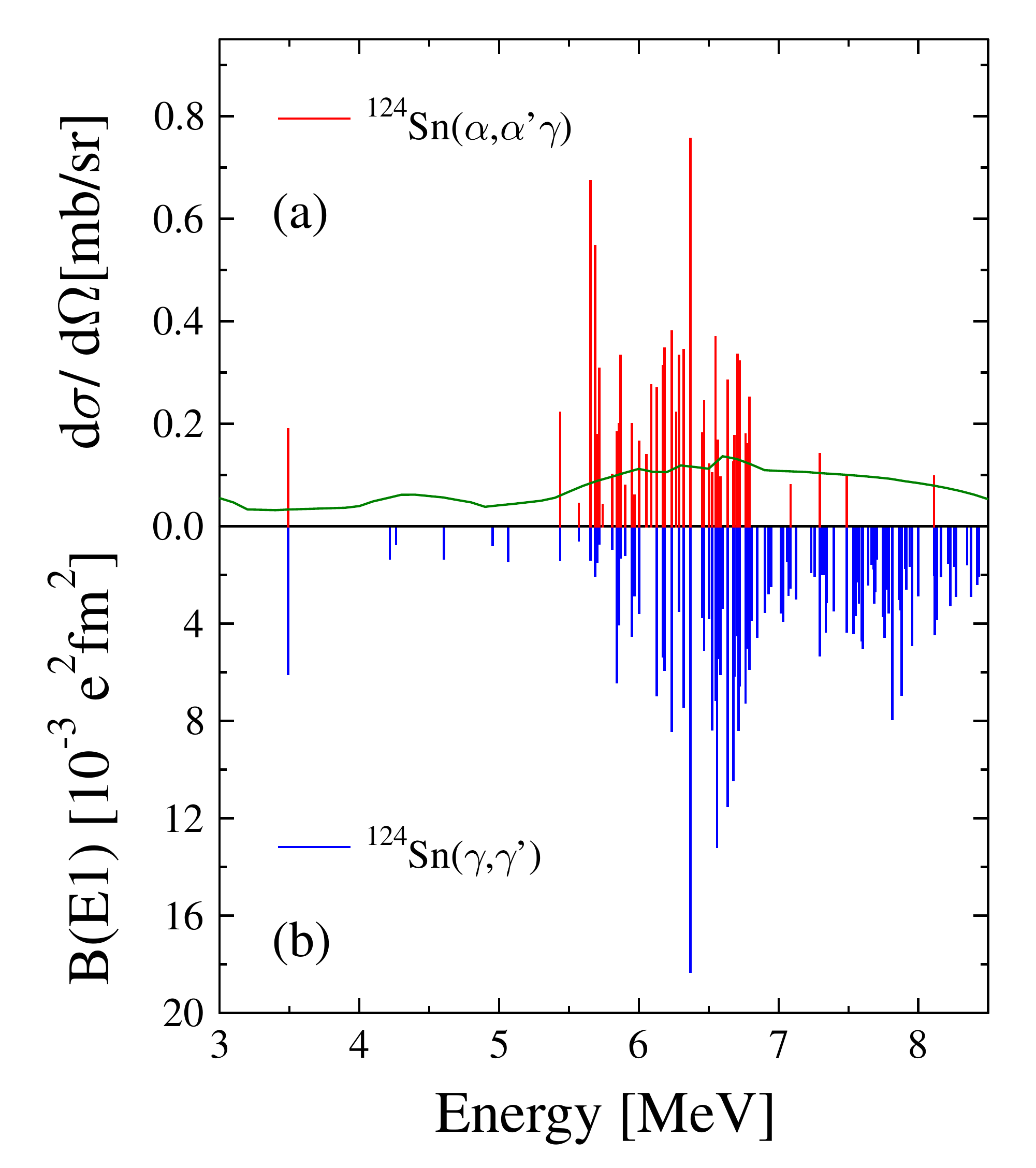}
\caption{(Color online)  Cross section for the excitation of the dipole states in $^{124}$Sn for the reaction ($\alpha, \alpha^{\prime} \gamma$) at $E_{\alpha}$=136 MeV  (panel a) is compared with the B(E1) measured with a ($\gamma,\gamma^{\prime}$) reaction (panel b). Taken with permission from \cite{End10}. \copyright 2010 by APS.}
\label{fig-end10}
\end{center}
\end{figure}
The cross section for the excitation of the dipole states in $^{124}$Sn - below the neutron separation energy - for the reaction ($\alpha,\alpha^{\prime} \gamma$) at $E_{\alpha}$=136 MeV \cite{End10} is shown in panel (a) of Fig. \ref{fig-end10}. The coincidence method is very selective and for many measured dipole states there is a one-to-one correspondence with the dipole states from ($\gamma, \gamma^{\prime}$) reactions shown in panel (b). However, there is an energy region where the dipole states are excited only by the electromagnetic field. This splitting of the PDR or isospin splitting seems to be an ubiquitous property of the PDR below the neutron emission threshold and it is still lacking a full understanding of the process. In fact, it has been suggested that the high-energy part of the PDR could correspond to the tail of the IVGDR but this is somehow in contradiction with the presence of PDR above the neutron separation energy for unstable nuclei. An attempt to find whether also for exotic nuclei the PDR splitting is present was done for the low-lying dipole states in $^{68}$Ni excited by an isoscalar probe such as $^{12}$C. The measurement done at LNS-INFN in Catania \cite{Mar18} has indeed shown that the low-lying dipole states of $^{68}$Ni have an isospin mixing typical of the PDR states. Unfortunately, due to the relatively low energy resolution no definite answer could be given regarding the presence of the isospin splitting. 

The isospin character of the low-lying dipole states has been studied using different types of isoscalar probes like $^{17}$O or protons. In all the nuclei investigated, the splitting of the PDR was confirmed (see the review \cite{Bra15} and references therein) establishing it as a general property of the PDR. The experimental data obtained with proton scattering at high energy (up to  80 MeV) and forward angles close to $0^{\circ}$ present an alternative method to investigate the pygmy resonance around the neutron emission threshold. Looking at the same nucleus with different probes, the results obtained with the proton beam show a stronger E1 strength compared to those obtained with the $\gamma$ beam. This is something that needs to be clarified with further experimental and theoretical studies.

The experimental evidences of the presence of the PDR can be summarised as follow. They are dipole states lying at an energy well below the IVGDR and with a few units of EWSR. They can be found only in neutron excess stable and unstable nuclei at energies above and below the neutron emission threshold. They can be excited by both isoscalar and isovector probes due to their strong isospin mixing. Below the neutron separation energy, the PDR states are separated in two parts: The dipole states belonging to the low-energy part are excited by both the electromagnetic and nuclear probes while in the higher energy region the states are populated only by the isovector interaction. This characteristic has been called PDR (or isospin) splitting and it is common to all the nuclei investigated until now. There are still some points that need to be clarified starting from the interplay between the isoscalar and isovector responses going to the collectivity (or not) of the PDR. Some recent experiments are planned to give  answers to this problem and try to disentangle the theoretical approaches devoted to this question. Another interesting aspect, which is worthwhile to investigate, is the presence of the PDR in strongly deformed nuclei. The very few experimental data available for deformed nuclei are not enough to give definitive answer to this problem. Some of the theoretical calculations reach different conclusions regarding the enhancement or the depletion of the PDR in deformed nuclei. 
 
\section{\textit{Theoretical approaches}\label{theo}}

\subsection{\normalsize{Macroscopic approaches}\label{theo-macro}}

Collective macroscopic models have been traditionally the pulling horses for the description of giant resonances, starting with the Isovector Giant Dipole Resonance \cite{Har01,Sat83}. The seminal approaches, based on liquid-drop model and hydrodynamical equations, are due to Goldhaber-Teller (GT) 
%\cite{Gol48} 
and Steinwedel-Jensen (SJ).
%\cite{Ste50}. 
In the former case the two proton and neutron densities ${\rho}_p(r)$ and ${\rho}_n(r)$ are treated as incompressible fluids and the resonance arises from the oscillation of the proton sphere against the neutron one. In the latter approach only the total density is incompressible and the movement associated with the dipole mode is due to oscillations of the proton and neutron fluids back and forth inside the rigid sphere. The two models lead to different estimates for the mass dependence of the energy of the resonance ($A^{1/6}$ for GT and $A^{1/3}$ for SJ), but more interestingly lead to different radial dependence of the corresponding transition densities induced by the ``isovector'' operator \cite{Cat97a}
\bea
\delta \rho^{(SJ)}_{iv} (r) &=& \alpha_{1} \Big( \frac{2NZ}{A}\Big) \> r \rho(r) \\
\delta \rho^{(GT)}_{iv} (r) &=& \delta \rho_{n} -\delta \rho_{p} =   \beta_{1}  
\Big[ \frac{2N}{A} \frac{d}{dr} \rho_{p}(r) + \frac{2Z}{A} \frac{d}{dr} \rho_{n}(r)\Big]
\eea
where the parameters $\alpha_{1}$ and $\beta_{1}$ are fixed to reproduce the total B(E1) value.  Note that, as apparent for example from the transition density in the GT model, if the proton and neutron densities have a similar radial profile (as it is expected for systems in the stability valley and $N\approx Z$) the corresponding ``isoscalar'' transition density
\be
\delta \rho^{(GT)}_{is} (r) = \delta \rho_{n} +\delta \rho_{p} = \beta_{1}  
\Big[ \frac{2N}{A} \frac{d}{dr} \rho_{p}(r) - \frac{2Z}{A} \frac{d}{dr} \rho_{n}(r)\Big]
\ee
vanishes. This amounts to say that the nature of the IVGDR is purely isovector and, for example, cannot be excited by isoscalar probes such as inelastic ($\alpha$,$\alpha^{\prime}$) scattering. As it is known, moving out from the stability valley, the neutron-rich systems start to display a neutron density that extends to larger radii.  As a consequence of this ``neutron skin'' there is a non-vanishing  isoscalar transition density (in leading order proportional to the size of the neutron skin) and the IVGDR acquires a mixed isovector/isoscalar character. It should finally be recalled the presence of the compressional Isoscalar Giant Dipole Resonance (ISGDR), at higher excitation energy than the IVGDR, generated by the operator $\sum_{i} r_{i}^{3} Y_{10} (r_{i})$, which is the leading non-spurious term in the expansion of $j_{1} (qr) Y_{10}(r)$ in the electromagnetic field. Assuming that the isoscalar dipole energy-weighted sum rule is fully exhausted by a single collective state, the corresponding isoscalar transition density can then be derived \cite{Har81}
\be \label{tdhd}
\delta \rho^{ISGDR} (r) = - \beta \big[3r^2 {d \over dr }+10r-
{5 \over 3 } <r^2> {d \over dr } + 
\epsilon (r {d^2 \over dr^2 } +4 {d \over dr })\big] \rho_{_0} (r)
\ee
with a node in the interior of the nucleus like the monopole breathing mode. The parameters $\beta$ and $\epsilon$ depend on the dimension of the nucleus.
\begin{figure}[!htb]
\begin{center}
\includegraphics[width=8cm, angle=0]{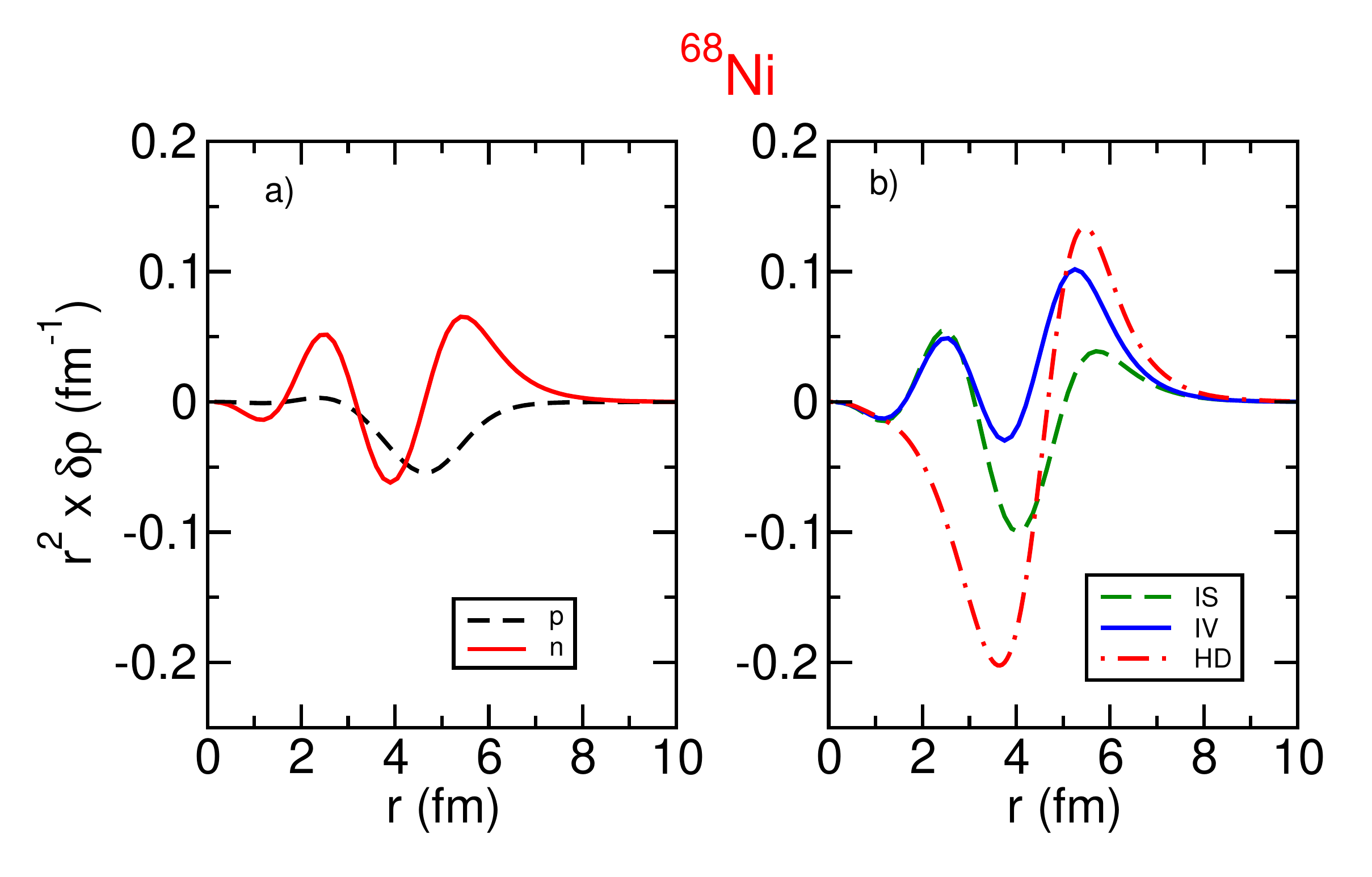}
\caption{ In panel a), proton (dashed black line) and neutron (solid red line) transition densities obtained within the macroscopic model, Eqs.  (\ref{tdour-p}) and (\ref{tdour-n}), respectively, for $^{68}$Ni. In panel b), the isovector (solid blue line) (Eq. \ref{tdour-iv}) and isoscalar (dashed green line) (Eq. \ref{tdour-is}) transition densities in comparison with the isoscalar transition density (HD) (dot-dashed red line) as deduced for the ISGDR in Ref. \cite{Har81}, Eq. (\ref{tdhd}). All the transition densities have been normalised to the microscopic RPA values.}
\label{td-macro}
\end{center}
\end{figure}

Guided by the successful description of the IVGDR in terms of macroscopic models, similar approaches have been used also for the description of the Pygmy Dipole Resonance. The clear experimental evidence of the connection between the occurrence of the PDR and the presence of a significant neutron excess leads in a natural way to the dynamical interplay of three actors, namely the proton and neutron cores plus the external valence neutrons of the skin. In the generalisation of the SJ model \cite{Suz90} the three incompressible oscillating fluids are all confined within the same sphere.  The key role played by the neutron skin and the importance of a different radial distribution for the core and the valence terms makes more attractive and successful the generalisation of the GT model \cite{Van92}.  More precisely the total density is divided into
\be
{\rho (r)} = \rho_{_p} (r) + \rho_{_n}^C (r) + \rho_{_n}^V (r) 
\ee
in terms of proton, neutron core ($N_{C}$) and neutron valence ($N_{V}$) densities.  The PDR mode is assumed to be associated with the dipole out-of-phase oscillation of the neutron valence density with respect to the core, given by the combined proton and neutron (core) densities.  One can immediately realise the mixed isoscalar/isovector nature of this mode, as apparent from the corresponding transition densities as given in Ref. \cite{Lan15}
\bea
\label{tdour-isiv} 
{\delta \rho_{is}}&=&\delta \left[ {N_V \over A}~{d(\rho_{_n}^C + \rho_{_p})\over dr}-{N_C+Z \over A}
{d\rho_{_n}^V \over dr} \right] \label{tdour-is} \\ 
{\delta \rho_{iv}} &=& \delta \left[ {N_V \over A}~{d(\rho_{_n}^C - \rho_{_p}) \over dr} -{N_C+Z \over A}
{d\rho_{_n}^V \over dr} \right]
\label{tdour-iv}
\eea
that are shown in panel b) of Fig. \ref{td-macro}, together with the separate individual proton and neutron contributions (panel a) for the isotope $^{68}$Ni
\bea
\label{tdour} 
{\delta \rho_{_n}(r)} &=& \delta \left[ {N_V \over A}~{d\rho_{_n}^C (r) \over dr}-{N_C+Z \over A}
{d\rho_{_n}^V (r) \over dr} \right] \label{tdour-n} \\
{\delta \rho_{_p}(r)} &=& \delta \left[ {N_V \over A}~{d\rho_{_p} (r) \over dr}  \right]
\label{tdour-p}
\eea
where $\delta$ is the deformation length.
Clearly protons and neutrons are oscillating in phase in the interior of the system, at variance with the external region dominated by the out-of-phase neutron contribution. Note that the complete dominance of the neutron component in the tail region mixes the isoscalar and the isovector characters of the mode in the external region, with obvious consequences for those reactions that are more sensitive to the tail of the systems (see next sections). This is more evident when a comparison is made with the isoscalar transition density of Ref. \cite{Har81}, Eq. (\ref{tdhd}), which is shown in panel b). 
The ratio of the PDR energy relative to the IVGDR one can be expanded in power of the ``neutron skin'' $y=R_{n}-R_{p}$ and eventually can be expressed as function of the average radius $\bar{R} = (R_{n}+R_{p})/2$ and $y$ \cite{Van92}
\be
\frac{E_{PDR}}{E_{IVGDR}}  = C \> \Big(\frac{Z}{Z+N_{V}}\Big)^{1/2}\> \Bigg[ 1-\frac{\bar{R} \> y}{20 \> a^{2}} \Bigg]
\ee
where $a$ is the diffusenesses of the neutron and proton ground-state densities which have been assumed equal. The constant C contains terms in ${\bar R}^{2}$ and $a^{2}$ \cite{Van92}.

\subsection{\normalsize{Microscopic approaches}\label{theo-micro}}

Although the macroscopic models give a physical insight of the collective or less collective modes, the mean-field  and the energy-density functional theories - based on our knowledge of the microscopic nature of the nucleus -  provide an accurate and deep understanding of such phenomena. These theories have successfully described all the main characteristics of the Giant Resonances as their energies, their strengths and the damping mechanism that generates theirs widths. The basic formulation and the more sophisticated implementations have been applied also to study the low-lying dipole states in nuclei with neutron excess. The successful description of the main properties of the PDR, even for nuclei far from the stability line, testifies to the robust structure of these theories. The detailed description of these theories is given in specific books \cite{Row10,Rin04} or recent reviews \cite{Paa07,Roc18,Lan22}. In the following a brief overview is given quoting the main aspects and results regarding this study.

Starting from an effective nucleon-nucleon interaction, the potential generated by all the nucleons in a nucleus is constructed with the Hartree-Fock (HF) method, which consists in solving the Schr\"odinger equation iteratively until self-consistency is obtained \cite{Rin04}. The frequently used effective nucleon-nucleon interactions are the zero-range Skyrme interaction \cite{Sky58} or the finite-range Gogny interaction  \cite{Dec80}. Both of them have several parameters that are fixed to reproduce the main properties of the ground states of nuclei. They are able to reproduce - with great accuracy - also the correct order of the single-particle levels for closed-shell nuclei. The residual interaction - the difference between the two-body interaction and the mean field obtained by the H-F method - is responsible for the elementary excitation of one-particle one-hole ($1p-1h$) configuration, which corresponds to the promotion of a particle above the Fermi level leaving a hole in the level below. The residual interaction is also responsible for the coherent mixing of $p-h$ configurations, with the same angular momentum, giving rise to a collective state. Such excitations are described by the Random-Phase Approximation (RPA) whose equations can be deduced by the equation-of-motion method \cite{Row10,Rin04}. The RPA equations are written in compact form as 
\be
\label{rpa-eq}
\left(
\begin{array}{cc}
  A &  B \\
  B^* &  A^* 
\end{array}
\right)
\left(
\begin{array}{c}
X^\nu \\
Y^\nu
\end{array}
\right)
= \hbar E_\nu
\left(
\begin{array}{c}
X^\nu \\
-Y^\nu
\end{array}
\right)
\ee
where the matrix $A$ and $B$ are defined in terms of the commutators of the Hamiltonian $H$ with the $ph$ creation and annihilation operators
\be
A_{p h p^\prime h^\prime} = <HF|[a_h^\dagger a_p [H, a^\dagger_{p^\prime} a_{h^\prime}]|HF>
\ee
\be
B_{p h p^\prime h^\prime} = -<HF|[a_h^\dagger a_p [H, a^\dagger_{h^\prime} a_{p^\prime}]|HF> \>.
\ee
The expectation values are calculated using the ground state $|HF>$ instead of the correlated ground states $|RPA>$ under the assumption that they are not much different. This is called quasi-boson approximation because it is assumed that the $ph$ creation or annihilation operators behave as boson operators. An excited collective state $|\Psi_\nu>$ is described as superpositions of $p-h$ and $h-p$ configurations with respect to the correlated ground state $|\Psi_0>$. The operator that creates such state can be written as 
\be
q^\dagger_\nu = \sum_{ph} \big[ X^\nu_{ph}a^\dagger_p a_h - Y^\nu_{ph} a^\dagger_h a_p \big] 
\label{q-oper}
\ee
where the amplitudes X and Y are solutions of the RPA secular equation. The ground state is defined as the vacuum of the $q_\nu$ operator $q_\nu |\Psi_0> =0$ and $|\Psi_\nu> = q^\dagger_\nu |\Psi_0>$. Since these $p-h$ amplitudes can be written as 
\bea
X^\nu_{ph} = <\nu |a^\dagger_p a_h |0> \\ 
Y^\nu_{ph} = <\nu |a^\dagger_h a_p |0>,
\eea
their absolute square gives the probability to find the configuration $a^\dagger_p a_h |0>$ and $a^\dagger_h a_p |0>$ in the excited state $\nu$. The term $Y$ describes the correlation in the ground state and when these amplitudes are zero the RPA equations reduce to what is known as the Tamm-Dancoff Approximation. The RPA solutions are superposition of many p-h configurations and when they sum up coherently they correspond to collective states. The  low-lying or giant resonances collective vibrational states are very well described by this method and their excitation is calculated using electromagnetic (isovector) or hadronic (isoscalar) operators which in the case of dipole states, and in the long wavelength limit, are given by
\be
O^{(IV)}_{1M}= {eN\over A}  \sum_{p=1}^Z r_p Y_{1M}(\hat r_p) - 
{eZ\over A} \sum_{n=1}^N r_n Y_{1M}(\hat r_n)  .
\label{oiv}
\ee
where the effective charges for protons (${eN\over A}$) and neutrons (${eZ\over A}$) have been introduced to remove the centre-of-mass motion, which is a spurious translational mode. For the isoscalar dipole operator, the lowest term of the expansion corresponds to a spurious translational motion and therefore the next-order term, which corresponds to a $3\hbar \omega$ dipole nuclear transition, is considered and the leading-order dipole transition operator can be written as
\be
O^{(IS)}_{1M}= \sum_{i=1}^A (r_i^3 - {5 \over 3}<r^2> r_i)Y_{1M}(\hat r_i) 
\label{ois}
\ee
where the term (${5 \over 3}<r^2>$) has been introduced to eliminate the spurious contribution of the centre of mass. The response to the excitation operator is given in terms of the reduced transition probability from the ground state to the excited state $\nu$ and it can be written as
\be
B(E\lambda, 0 \rightarrow  \nu)  =\Bigg|\sum_{ph} (X_{ph}^\nu - Y_{ph}^\nu)
 <p|| O_\lambda ||h> \Bigg|^2  = \Bigg|\sum_{ph} b_{ph}(E\lambda) \Bigg|^2
 \label{bel}
\ee
where $<p|| O_\lambda ||h>$ are the reduced multipole transition amplitudes associated with the elementary $p-h$ configurations. The $b_{ph}(E\lambda)$ are the partial contribution of a $p-h$ configuration to the reduced transition probability for the given state.

A quantity that contains much information on of the excited state is the transition density which is defined in terms of the off-diagonal matrix element of the ground state density. Its radial part gives information on where the excitation is localised (volume or surface) or on the isoscalar or isovector nature of the excitation, namely whether the neutron-proton motion is in- or out-of-phase. The transition density for a state $\nu$ with an angular momentum $\lambda$ in the RPA approach can be written as 
\bea
\nonumber
\delta \rho^\nu &=& {1 \over \sqrt{4 \pi}} \sum_{ph} {\hat j_p \hat j_h \over \hat \lambda}
(-)^{j_p+j_h- {1 \over 2} } <j_h {_1 \over ^2} \, j_p -{_1 \over ^2} |\lambda 0> \times \\
&&\delta (\lambda+l_p+l_h, even) [X^\nu_{ph} - Y^\nu_{ph}] \, R_{l_p j_p} (r) R_{l_h j_h} (r)
 \label{td}
\eea
where the $X$ and $Y$ are the RPA amplitudes, $\hat\lambda= 2\lambda +1$, $l$ and $j$ are the orbital and the total angular momenta of the single-particle states, respectively. Their radial wave functions $R$ are solutions of the HF equations. The transition densities for protons and neutrons can be calculated by running the summation separately over the number of $p-h$ configurations of protons and neutrons, respectively. The isoscalar and isovector transition densities can be constructed as 
\bea
\delta \rho^\nu_{IV} (r) = \delta \rho^\nu_n(r) - \delta \rho^\nu_p(r)  \\
\delta \rho^\nu_{IS} (r) = \delta \rho^\nu_n(r) + \delta \rho^\nu_p(r)
 \label{td-isiv}
\eea
\begin{figure}[!htb]
\begin{center}
\includegraphics[width=6cm, angle=0]{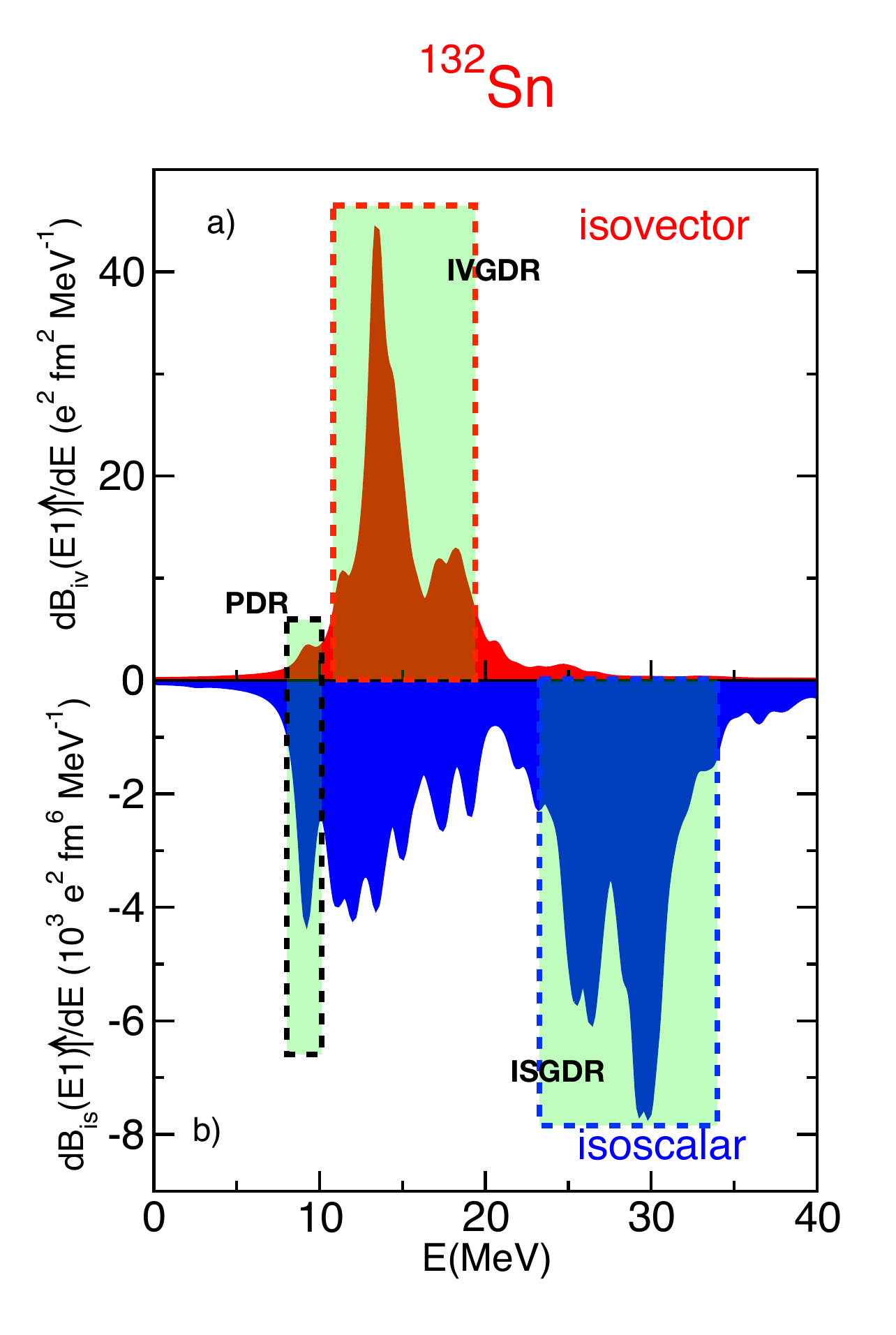}
\caption{(Color online)  Isovector (panel a) and isoscalar (panel b) RPA dipole strength distributions in $^{132}$Sn. The highlighted regions correspond to PDR, IVGDR and ISGDR modes.}
\label{fig-belsn}
\end{center}
\end{figure}

\begin{figure}[!htb]
\begin{center}
\includegraphics[width=8cm, angle=0]{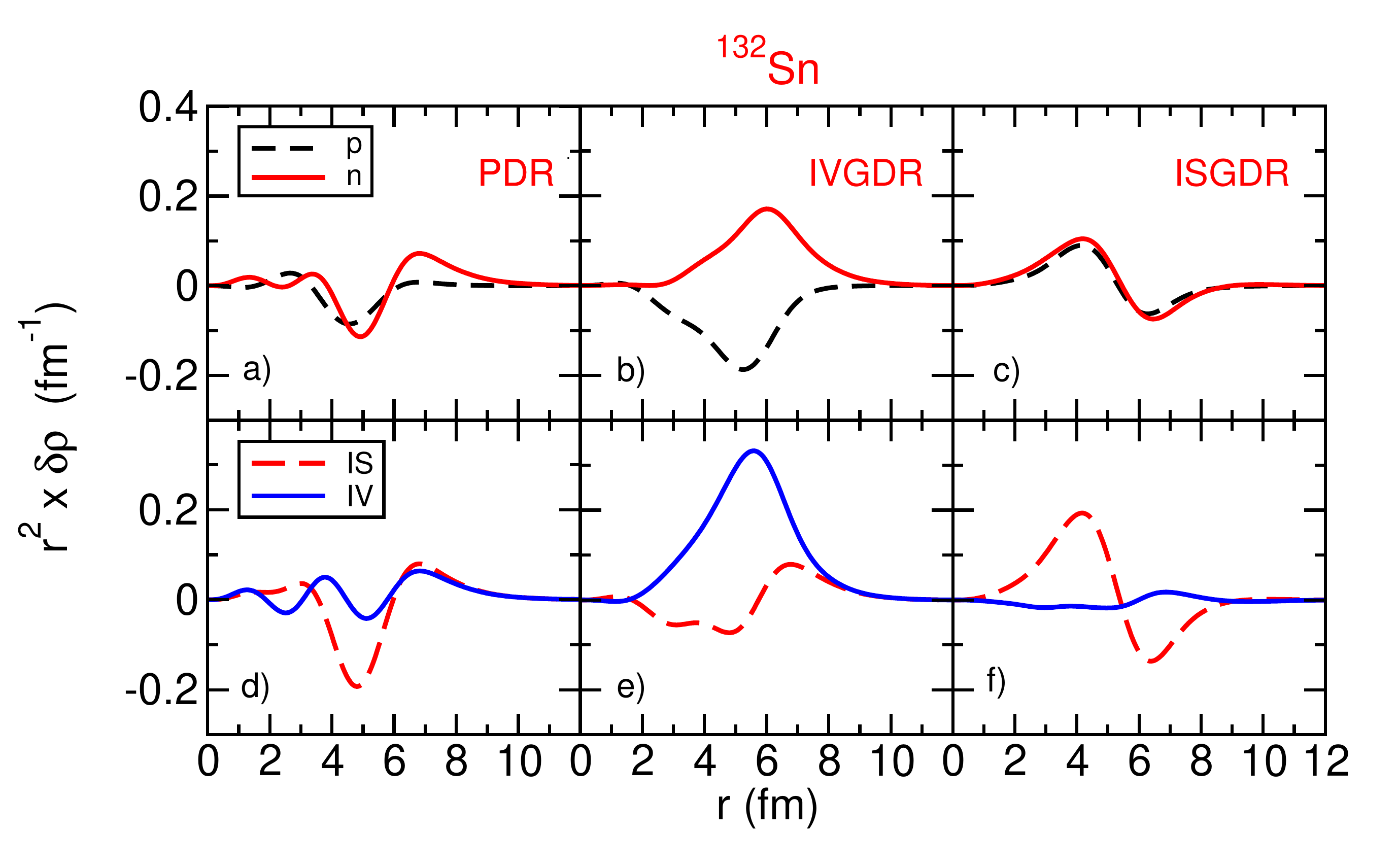}
\caption{(Color online)  In the top panels, the RPA proton (dashed black line) and neutron (solid red line) transition densities are shown for the three shaded areas of the dipole strength distributions in $^{132}$Sn of Fig. \ref{fig-belsn}. The corresponding isovector (blue solid line) and isoscalar (red dashed line) transition densities are shown in the bottom panels. }
\label{fig-tdsn}
\end{center}
\end{figure}

Early calculations based on HF plus RPA were performed to study the effects of the neutron excess on collective states in nuclei far from the stability line. The dipole strength distributions show a spreading of the strength and a shift towards lower energy with increasing neutron number. The isovector distribution presents a small peak at energies lower than the IVGDR as the neutron number increases while in the isoscalar response the low-energy peak - present also for isotopes with $N=Z$ - becomes more intense \cite{Ham96, Ham96a, Cat97a, Cat97b, Ham98}. Calculations performed within the HF + RPA framework to determine the dipole strength distribution for the two experimentally investigated exotic nuclei, i.e. $^{132}$Sn and $^{68}$Ni, reproduce the PDR low-lying peak. As an example, the isovector and isoscalar dipole reduced transition probabilities are shown in panels a) and b) of Fig. \ref{fig-belsn}, respectively. The calculations are done with a discrete HF plus RPA with a SGII Skyrme interaction \cite{Gia81,Gia81a}. The discrete dipole states obtained are convoluted with a 1-MeV width Lorentzians to produce the continuous curves shown in the figures. The isovector response in panel a) is generated by the operator of Eq. (\ref{oiv}) while the isoscalar  $3 \hbar \omega$ dipole transition is given by the operator in Eq. (\ref{ois}). The shaded areas indicate the three different dipole modes. In fact, the three different responses of the same nucleus to an isovector or isoscalar probe are reflected in the shape and magnitude of the corresponding transition densities, which are shown in Fig. \ref{fig-tdsn} for the representative states of the three energy regions.
It is evident, by looking at the transition densities, that the three highlighted regions represent different excitation modes. For the IVGDR mode, around 15 MeV, the proton and neutron transition densities (panel b) are out of phase - in accordance with the macroscopic models - and therefore this mode is almost pure isovector as is clearly seen from the isovector transition density in panel e). On the contrary, the transition densities for the ISGDR around 30 MeV are in phase inside the nucleus and at the nuclear surface producing an isoscalar mode whose transition density (panel f) has the same shape of a compressional mode. For the PDR state, the proton and neutron transition densities are in phase inside the nucleus while at the surface only the neutrons give the surviving contribution. Therefore, for this excitation at the nuclear surface, the isoscalar and isovector transition densities have the same shape and strength. In the case of an excitation process to populate the PDR states with medium-heavy ions, the explored region is mainly the surface of the target nuclei. As a consequence, this new mode can be explored experimentally by both isovector and isoscalar probes. This has been exploited and as a consequence the new feature of the splitting of the PDR has been found.

This feature of the PDR can be considered as a general characteristic of this mode in the sense that it has been found in all the nuclei with neutron excess. Furthermore, all the microscopic theories (briefly described below) give the same general description of the transition densities even though they might differ for some specific aspects of the PDR. This can be considered as a kind of theoretical definition of the Pygmy Dipole Resonances.

The sharp separation between the occupied and unoccupied levels below and above the Fermi level is not satisfied in open-shell nuclei. For these nuclei the pair interaction is very important and should be included in the description of the ground-state properties. This can be achieved within the Bardeen, Cooper and Schrieffer (BCS) model or in its self-consistent version, the Hartree-Fock-Bogoliubov (HFB) theory which provides the basis space for the so-called quasiparticle RPA (QRPA) \cite{Rin04}. In this approach the operator responsible for the excitation, in analogy with Eq.~(\ref{q-oper}), is
\be
Q^\dagger_\nu = \sum_{k k'} \big[ X^\nu_{k k'}\alpha^\dagger_k \alpha^\dagger_{k'} - Y^\nu_{k' k} \alpha_{k'} \alpha_k \big] \>.
\label{q-oper-2}
\ee 
where the Bogoliubov transformation has been used
\bea
\alpha^\dagger_k = u_k a^\dagger_k - v_k a_{-k} \\
\alpha^\dagger_{-k} = u_k a^\dagger_{-k} + v_k a_{k}
\eea
with $|-k>$ the time reversal state of $|k>$ and $u_k^2 + v_k^2 = 1$. This approach is suited not only for the open-shell nuclei but also for the deformed ones too. A complete and detailed description can be found in Ref. \cite{Rin04}. 

QRPA calculations with few different Skyrme effective interactions were performed for even Ca, Ni and Sn isotopes  from the proton to the neutron drip line \cite{Ter06}. Other systematic studies on even Ca, Sn and Ni isotopes based on a self-consistent QRPA were carried out \cite{Pap12, Pap14}, with a Gogny DS1 \cite{Ber91} finite-range interaction. In all these calculations, the strength found for the isotopes with $N=Z$ is shifted to lower energies with increasing neutron number. The calculated transition densities are pure isoscalar for $N=Z$; for neutron-rich nuclei, the contribution of the neutrons at the nuclear surface becomes predominant. These results are similar to the ones obtained with the HF + RPA for closed-shell nuclei and do not change much when using different interactions.

The time-dependent relativistic mean-field model developed in the seminal work of Walecka \cite{Wal74} was applied to describe the dynamics of collective motion \cite{Vre95}. The nucleus is described as a system of Dirac spinors and their interaction is mediated by the exchange of virtual mesons ($\sigma$-, $\omega$- and  $\rho$-mesons) and photons. The coupled equations of motion are given by the Dirac equation for the nucleons and by the Klein-Gordon equation for the mesons. The Relativistic Random-Phase Approximation (RRPA) represents the small amplitude limit of the time-dependent relativistic mean-field theory. It can be obtained by the linear response of the density matrix to an external field \cite{Nik02}. Several nuclear structure phenomena are well described by the RRPA and special attention was given to the PDR \cite{Vre01,Pie06,Lia07,Pie11,Vre12}. Note that the pairing correlation needs to be included for open-shell nuclei. The Relativistic Quasi-particle Random-Phase Approximation (RQRPA) can be derived, in the limit of small oscillations, from the Relativistic Hartree-Bogoliubov theory. A solution of the RQRPA equations is given in  Ref.~\cite{Paa03} by writing the relativistic Hartree-Bogoliubov wave functions in terms of BCS-like wave functions. This approach has been used to study isovector dipole strength in the Sn isotopes, where a NL3 effective interaction \cite{Lal97} for the relativistic mean-field of the effective Lagrangian and a Gogny D1S \cite{Ber91} for the phenomenological pairing interaction were used. The results show the appearance of a peak in the low-energy region whose strength is increasing with the neutron number while the centroid position shifts towards lower energies. The inclusion of the pairing interaction moves the position of the peak slightly to lower energies.

The RPA and QRPA are able to describe all the main properties of the Giant Resonances except their width generated by the damping mechanism. This is responsible for the spreading width given by the coupling of the $1p-1h$ configuration with states formed by $2p-2h$, $3p-3h$ or more complex configurations. There are several ways to implement such coupling, the most direct is the so-called Second RPA (SRPA) where the $2p-2h$ configurations are explicitly coupled with the $1p-1h$ configurations. In this extension, the excitation operator of Eq. (\ref{q-oper}) contains now a superposition of $1p-1h$ and $2p-2h$ configurations
\be
q^\dagger_\nu = \sum_{ph} \big[ X^\nu_{ph}a^\dagger_p a_h - Y^\nu_{ph} a^\dagger_h a_p \big] \>
+ \sum_{p<p^\prime,h<h^\prime} 
\big[ X^\nu_{php^\prime h^\prime}a^\dagger_p a_h a^\dagger_{p^\prime} a_{h^\prime} - 
Y^\nu_{php^\prime h^\prime} a^\dagger_h a_p a^\dagger_{h^\prime} a_{p^\prime} \big] \>.
\label{q-srpa}
\ee
Full SRPA calculations \cite{Pap09,Gam11} have been performed for O and Ca isotopes. The position of the PDR is better reproduced with respect to the RPA and RRPA results but the giant resonances are shifted to lower energies by several MeV in disagreement with the experimental values. This is ascribed to some other correlations, which are implicitly added by the SRPA procedure in the ground state. The procedure to determine the parameters of the effective interactions to reproduce the fundamental properties of nuclei include part of the complex configurations, which are explicitly introduced in the extended RPA. This double counting is adjusted by means of the so-called ``subtraction" model \cite{Tse07,Tse13}. An application of the Subtracted Second RPA (SSRPA) model, on nuclei with neutron excess to study the low-lying dipole states, can be found in Refs.~\cite{Gam18,Gam20} where the calculations have been performed using a SGII Skyrme interaction. The results for $^{68}$Ni seem to indicate the presence of the PDR splitting whose experimental evidence is still not clear because the only data available \cite{Mar18} do not clarify this aspect.
\begin{figure}[!htb]
\begin{center}
\includegraphics[width=6cm, angle=0]{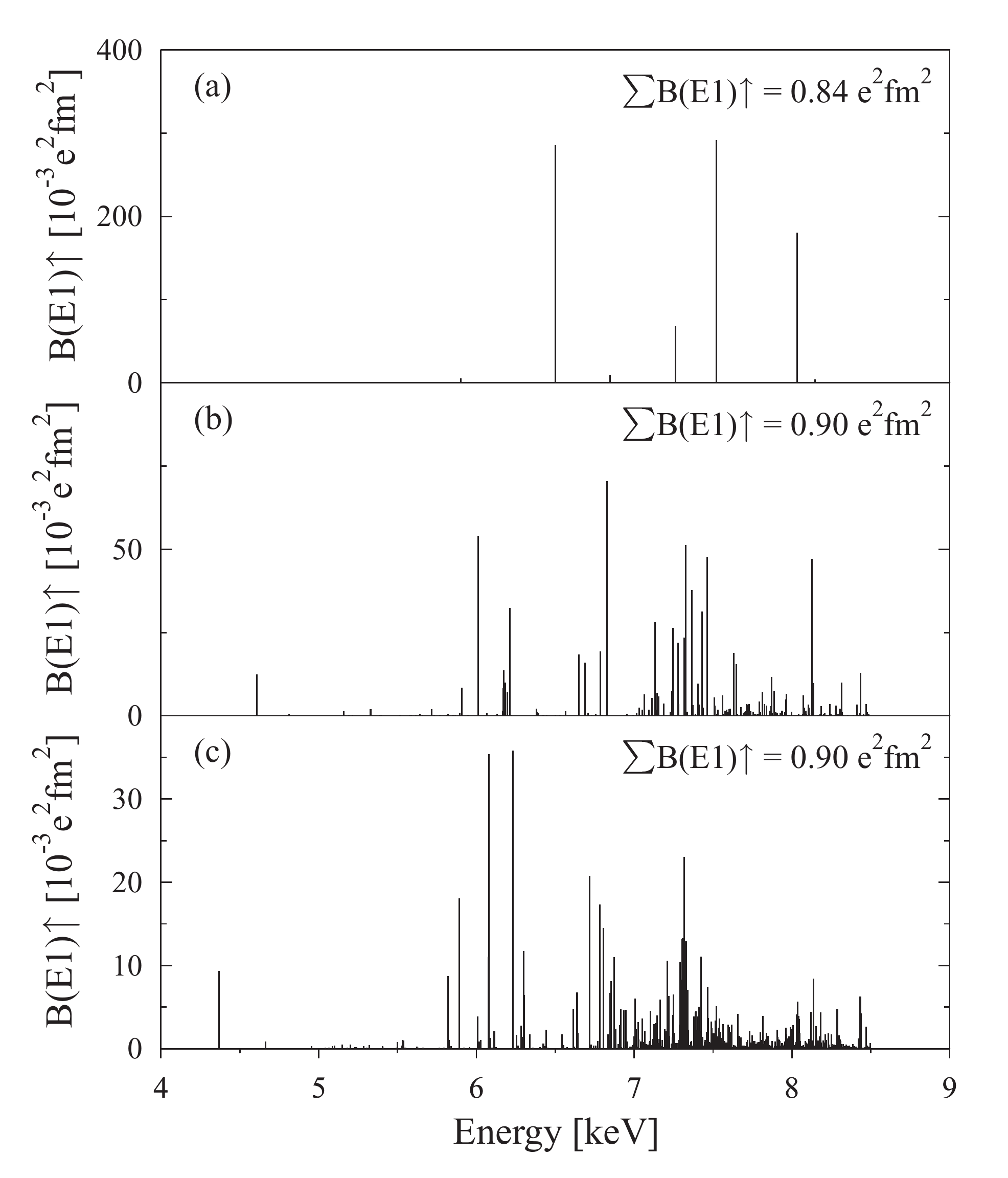}
\caption{ Dipole reduced transition probabilities B(E1) calculated within QPM for $^{136}$Xe for several approximations. The results when only one-phonon states are taken into account are shown in the top frame. Calculations when also the two-phonons and the two- and three-phonons are considered are shown in the middle and lower frames, respectively. Taken with permission from \cite{Sav11}. \copyright 2011 APS.}
\label{qpm}
\end{center}
\end{figure}

Another way to take into account higher order configurations is to explicitly couple the $1p-1h$ configuration with two- or three-phonon states. Writing the RPA as the lowest order of a boson expansion \cite{Rin04} the coupling can be introduced by taking into account higher order terms of the expansion \cite{Lan97,Fal03,Lan06} containing terms of the residual interaction capable to mix contributions coming from multi-phonon states. An approach similar to this is the so-called Quasi-particle Phonon Model (QPM) \cite{Sol92,Ber99} that has been successfully applied to the description of the PDR below the neutron emission threshold. The solutions of the QRPA give the one-phonon basis, which is employed to construct the two- and three-phonon states to enlarge the original basis. The Hamiltonian of the system is then diagonalised in this enlarged basis and the eigenfunctions are mixed states whose components are of one-, two-, and three-phonon type:
\be
\label{mix-3} %\nonumber
|\Phi_\alpha>  = \sum_{\nu_1} c^\alpha_{\nu_1} |\nu_1> + 
\sum_{\nu_1\nu_2} c^\alpha_{\nu_1\nu_2} |\nu_1 \nu_2> \\ 
+ \sum_{\nu_1\nu_2\nu_3} c^\alpha_{\nu_1\nu_2\nu_3} 
|\nu_1\nu_2\nu_3>  \> .
\ee
In this approach, it is preferred to use phenomenological central and spin-orbit Woods-Saxon potential instead of a functional interaction as the Skyrme ones. In this way, the parameters that describe the nuclear ground-state properties are fixed with great accuracy. The excitations are calculated assuming for the residual interaction a sum of isoscalar and isovector separable multipole interactions. The coupling of one-phonon state with more complicated configurations produces a fragmentation of the dipole strength into many states towards lower energies. Detailed studies on Sn isotopes have been done in Ref. \cite{Tso04,Tso08}. An example of how the coupling to configurations of increasing complexity modifies the dipole strength distribution is shown in Fig. \ref{qpm} \cite{Sav11}. In the top panel, the B(E1) distribution for $^{136}$Xe is obtained including only the one-phonon states. These results - that should be equivalent to an RPA calculation - are strongly modified when the coupling with two- (middle panel) and three-phonon (bottom panel) are taken into account. The states are fragmented into hundreds of states, their centroid moves to lower energy while the summed strength in almost unchanged. These last results are in quantitative agreement with the experimental values \cite{Sav11}. This approach describes with good accuracy some of the main features of the PDR but loses the fully microscopic picture and the self-consistency.
\begin{figure}[!htb]
\begin{center}
\includegraphics[width=10cm, angle=0]{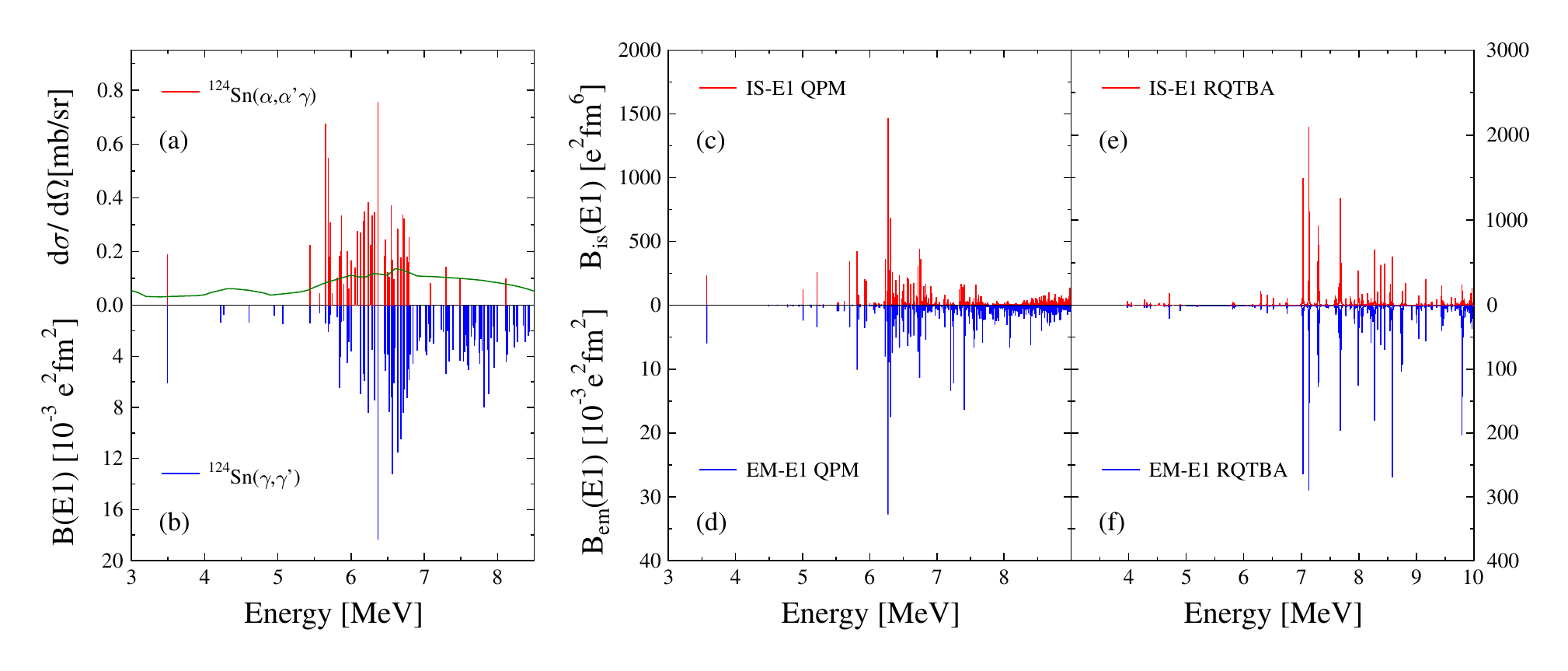}
\caption{(Color online)  Cross sections for the excitation of the dipole states in $^{124}$Sn in the reaction ($\alpha,\alpha^{\prime} \gamma$) at $E_{\alpha}$=136 MeV  (panel a) is compared with the B(E1) measured with a ($\gamma, \gamma^{\prime}$) reaction (panel b). The QPM (middle column) and RQTBA (right column) isoscalar $B_{is}(E1)$ (top panels) and electromagnetic $B_{em}(E1)$ (lower panels) are shown for comparison. Taken with permission from \cite{End10}. \copyright 2010 by APS.}
\label{fig-a-end10}
\end{center}
\end{figure}

These characteristics are recovered in the self-consistent Relativistic Quasiparticle Time-Blocking Approximations (RQTBA) where the RQRPA is modified by including the coupling to the low-lying vibrations. The Time-Blocking Approximation (TBA) - where the $p-h$ configurations are ordered in time - was modified to couple the quasi-particle states to collective degrees of freedom  \cite{Tse07,Lit07,Lit08}. The subtraction procedure \cite{Tse07} is included to avoid the double counting coming from the fitting - obtained with the relativistic mean-field - of the ground-state properties. The calculations within the RQTBA \cite{Lit08,Lit09}, done for several isotopes of Sn, Ni, and for some $N$=50 isotones, show a fragmented distribution of the dipole strength in the low-energy region maintaining the  IVGDR peak almost unchanged with respect to the RQRPA results.

The two approaches including the coupling with multi-phonon states describe reasonably well the isospin splitting of the PDR for $^{124}$Sn as shown in Fig. \ref{fig-a-end10}. The B(E1) measured with a ($\gamma, \gamma^{\prime}$) reaction (panel b) is compared with the electromagnetic $B_{em}(E1)$ calculated within the QPM (panel d) and RQTBA (panel f) showing a reasonable agreement with the general trend of the experimental data. In the top panels, the comparison is done between the dipole excitation cross section in $^{124}$Sn for the reaction ($\alpha,\alpha^{\prime} \gamma$) at $E_{\alpha}$=136 MeV (panel a) and the isoscalar $B_{is}(E1)$ calculated within the QPM (panel c) and RQTBA (panel e) approaches. While the direct comparison between Coulomb cross section and $B_{em}(E1)$ is correct - because they are proportional - the $\alpha$ scattering cross section cannot be directly compared with the calculated isoscalar $B_{is}(E1)$. In order to move the comparison to a quantitative level, inelastic cross section calculations have to be performed \cite{Lan14}. Details on such calculations will be treated in the next section.

It has been proposed that the PDR can be seen as a manifestation of the isoscalar Toroidal ~~Dipole ~~Resonance (TDR) \cite{Vre02, Kva11, Nes16,Rep19} - generated by a transverse oscillation of the nucleons - which contributes to the isoscalar dipole mode together with the compressional mode. It is estimated to be found at an excitation energy of $E_{tor} = (65-85) A^{-1/3}$. Numerical calculations have been done using fully self-consistent theories discussed above and its presence has been found for numerous stable and unstable nuclei independent from the neutron-to-proton ratio. Experimental evidences though show the PDR to be present in nuclei with $N>Z$ while the TDR is predicted in all nuclei. The TDR transition densities for the low-lying states show the same shapes described above for the PDR. However, this is not surprising since the transition densities do not depend on the toroidal operator \cite{Rep13} but rather on the RPA wave functions. The question of the presence of the TDR is under debate and selective experimental observations have yet to come.

Collective modes have been studied using the Vlasov equation that is the semi-classical limit of the Time-dependent Hartree-Fock (TDHF) approach. A detailed description of this method can be found in Ref. \cite{Urb12} where the Vlasov equation has been used to study the PDR in the O and Sn isotopes. A small enhancement in the isovector dipole strength function below the IVGDR centroid is present for the selected $N>Z$ nuclei while it is absent for nuclei with $N=Z$. With some approximations one can extract the velocity fields and the radial transition densities that have a PDR-like shapes.

Collective excitations have been described also within the Interacting Boson Model (IBM) \cite{Iac87}. Its extension by including the $p$ and $f$ bosons, besides the $s$ and $d$, was applied to calculate the dipole strength distribution of some $N$=82 isotones \cite{Pas12}. The IBM calculations are able to reproduce the shape of the distributions together with the slight increase of the centroid with isotone mass.

Recently, the study of the PDR in deformed nuclei has been carried out both experimentally and theoretically. In deformed nuclei with axial deformation, the IVGDR splits in two peaks that correspond - in the hydrodynamical model - to vibrations along the two principal axes. Considering the PDR as due to the oscillation of a neutron skin against an inert core then a splitting, due to the deformation, should also be expected for this strength. Until now experimental data \cite{God13} have not given any definite answer to this problem while some theoretical  calculations reach opposite conclusions. A relativistic Hartree-Bogoliubov (RHB) mean field plus relativistic QRPA microscopic calculations \cite{Pen08,Pen09} have been performed to investigate the PDR for several tin isotopes. The transition densities show a pattern similar to those found in spherical nuclei. On the other hand, the results regarding the summed B(E1) lead to the conclusion that deformation quenches the dipole response in the low-lying energy region. An opposite result is reached in a calculations performed within a HFB plus QRPA with Skyrme interactions \cite{Yos11} for Nd and Sm isotopes, where the summed low-lying dipole strength is found to be five times larger than in spherical nuclei. The effects of deformation on the dipole response is not yet well established and the development of additional investigations is important.

All the theoretical studies described above are devoted to the description of the structure of the PDR. They are centred on understanding and reproducing the strength distribution of the low-lying dipole states without losing the good description obtained for the IVGDR. When the investigation is done with isoscalar probes, such as $\alpha$ particle or $^{17}$O, the main measured quantities are the inelastic cross section and the $\gamma$-ray decay. In this case, the theoretical knowledge on the inelastic cross section can be very useful for the understanding and interpretation of the experimental data. In such process, the identity of the collision partners has to be preserved so that the reaction can be described as direct whose maximum cross section is found at forward (grazing ) angles, i.e. exploring the peripheral (surface) nuclear region. There are several methods to calculate the inelastic cross section. In the following section, some of the important aspects of those calculations will be presented.

\section{\textit{Cross-section calculation}\label{xsec-cal}}

Consider the inelastic scattering of a projectile $a$ impinging on a target $A$ (channel $\alpha$) going to a final outgoing channel $\beta$ corresponding to the outgoing nuclei $b$ and $B$. A set of Coupled-Channel (CC) equations - equivalent to the Schr\"odinger equation - can be deduced to describe the whole scattering process \cite{Sat83}.  These are an infinite set of coupled equations for all the possible internal states of the reaction partners. Due to the infinite number of channels involved in the scattering process a reduction to a finite number of physical important channels is necessary. The choice of the finite set of channels is often guided by the physical problem under investigation and/or the available experimental data. The effect of the discarded channels is taken into account by the choice of an appropriate optical potential. In the case in which there is only one strong final channel, the CC equations reduce to the so-called the Distorted-Wave Born Approximation (DWBA) which can be written in terms of the transition amplitude or T-matrix. The T-matrix gives the transition probability from the initial channel $\alpha$ to the final one $\beta$ induced by an interaction $V$  and it is related to the scattering amplitude $f_{\beta \alpha}$ as follow
\be
T_{\beta \alpha} = - \frac{2 \>\pi \>\hbar^{2}}{\mu_{\beta}}f_{\beta \alpha}
\ee
with $\mu_{\beta}$ the reduced mass of channel $\beta$. The transition probability for the reaction $A(a,b)B$ in the DWBA is
\be
T^{DWBA}_{\beta \alpha} = \int  \ \chi_\beta^{(-)}  (\vec k_{\beta} \vec r_\beta) < \psi_b \  \psi_B |  V |\ \psi_a \ \psi_A > \  \chi_\alpha^{(+)} (\vec k_{\alpha} \vec r_\alpha) \>d x
\label{tdwba}
\ee
the wave functions $\chi^{(\pm)}_{\alpha}$ describe the elastic scattering in the channels $\alpha$  due to the optical potentials $U_{\alpha}$. The plus and minus signs indicate the incoming and outgoing scattering wave functions, respectively. 

The validity of the DWBA lies in the fact that the nuclear potential is considered as the sum of two parts: one can be considered as a mean-field potential, which describes the collision between the two nuclei, and the other one as a residual interaction, relatively small, responsible for the excitation and to be considered as a perturbation. This approximation is valid when the elastic scattering is the most important event in a nuclear collision while the other reaction channels can be considered as perturbations. The DWBA is obtained in a first-order perturbation, which is equivalent to saying that the nuclear interaction is one-step process, i.e. the interaction acts only once. In the Coupled-Channel framework, the interaction is allowed to act many times producing a more realistic description of the nuclear reaction.

For medium/heavy reaction partners and incident energies close to the Coulomb barrier, a semi-classical description of the nuclear collision where the relative motion of the centre of mass of the ions obeys the Newton equation of motion and the excitation of the nuclei described according to quantum mechanics is justified \cite{Bro72,Bro04}. One implicit condition for the applicability of the semi-classical models is that the two colliding nuclei do not change their masses and charges during the nuclear process. This condition is satisfied in the case of the reactions used to study the PDR because these are direct reactions. To apply the semi-classical models the de Broglie wave length $\lambdabar$ should be small compared to the characteristic length of a nuclear reaction such as the distance of closest approach or the diffuseness of a Woods-Saxon potential, i.e. a distance where the potential changes significantly. In the case of Coulomb potential, a parameter can be defined from the atomic numbers $Z_{a}$ and $Z_{A}$ of the two colliding nuclei that have an asymptotically relative velocity $v$
\be
\eta = \frac{Z_a Z_A e^2}{\hbar v} \>.
\ee
Consider the expression of the distance of closest approach for impact parameter zero and the de Broglie wave length
 \be
d_0 = \frac{Z_a Z_A e^2}{\frac{1}{2}\mu v^2} \>\>; \>\>\> \lambdabar = \frac{\hbar}{\mu v} \>.
\ee
The parameter $\eta$ is then the ratio between half the distance of closest approach and $\lambdabar$. When $\eta \gg 1$, it is justified to consider that the nuclei move on classical trajectories.  When the nuclear potential is taken into account the relation is more complicated, but for medium and heavy ions colliding at energies above the Coulomb barrier the classical approximation is valid. 

Let us consider a projectile $a$ impinging on a target $A$ and assume that one can distinguish the reaction partners  along the entire classical trajectory. The solution of the scattering problem can be found solving the time dependent Schr\"odinger equation 
\be
i\> \hbar \frac{\partial}{\partial t} |\psi(t)> = H(t) |\psi (t)>
\ee
The Hamiltonian of the system is the sum of the Hamiltonians of the two partners of the reaction. 
\be
H (t) = H_a(t) + H_A(t)  ~~~~~~~~~~ {\rm where} ~~~~~~~~~ H_i(t) = H_i^0 + W_i (t) \> , ~~~{\rm with}\> i=a,A 
\ee
The Hamiltonian of the nuclei $a$ and $A$ can be written, in terms of creation and annihilation operators, as 
\be
H^0 = \sum_i \epsilon_i a^\dagger_i a_i + \sum_{ij,lk} V_{ij,lk} a^\dagger_i a^\dagger_j a_l a_k  \>.
\ee
The time dependence enters in the interaction term $W(t)$ through the relative distance $R(t)$ between the two nuclei
\be
W_A(t) = \sum_{ij} <i|U_a(R(t))|j> a^\dagger_i a_j + h.c.
\ee
it describes the excitation of the target $A$ due to the mean field $U_{a}$ of the projectile $a$. The eigenfunction of $H$ can be written as 
\be
|\Psi (t)> = |\psi_a (t)> |\psi_A (t)> \>
\ee
where the wave functions $\psi$ can be obtained by solving the Schr\"odinger equation (for the nucleus target $A$)
\be
i \> \hbar \frac{\partial |\psi_A(t)>}{\partial t} = H_A |\psi_A(t)>
\ee
with the initial condition that the nucleus is in the ground state $|\psi_A(-\infty)>  = |\psi_A^0>$. Expanding $|\psi_A(t)>$ over the eigenfunction $\phi_\alpha$ of $H^0$  ($H^0 |\phi_\alpha> = E_\alpha |\phi_\alpha>$) and omitting the indices $a$ and $A$,
\be
|\psi(t)> = \sum_\alpha a_\alpha(t) e^{-\frac{i}{\hbar} E_\alpha t} |\phi_\alpha> 
\ee
and substituting $|\psi(t)>$ in the Schr\"odinger equation, a coupled-channel equation for the amplitude $a_\alpha(t)$ is obtained
\be
\dot a_\alpha(b,t) = - \frac{i}{\hbar} \sum_{\alpha^\prime} e^{-\frac{i}{\hbar} 
(E_\alpha - E_{\alpha^\prime}) t} <\phi_\alpha|W(t)|\phi_{\alpha^\prime}> \> a_{\alpha^\prime} (b,t) \>.
\label{class-cc}
\ee
These coupled-channel equations have to be solved for each value of the impact parameter $b$ and with the initial condition that the nucleus before the collision should stay in its ground state, $a(-\infty) = \delta_{\alpha, gs}$. The use of the semi-classical coupled-channel equations, when justified, is more convenient than the quantum CC because the calculation can be guided by a physical insight and, more important, the number of channels included in the calculations can be orders of magnitude larger. 

The solution of the semi-classical CC provides the probability amplitude as a function of $b$. The value of these amplitude at the end of the scattering process ($+\infty$) gives the excitation probability of the state $\alpha$
\be
P_\alpha (b) = |a_\alpha (b, +\infty)|^2
\ee
The cross section for the inelastic excitation of the state $\alpha$ is obtained by integrating $P_\alpha (b)$ over the impact parameter range
\be
\sigma_\alpha = 2 \pi \int_0^{+ \infty} P_\alpha (b)\> T(b) \>b \> db \>\> .
\label{xsec}
\ee
Processes not explicitly included in the model space are taken into account by the transmission coefficient $T(b)$. These processes reduce the incident flux when other channels are opened. The coefficient is usually taken as a depletion factor that falls to zero as the overlap between the two nuclei increases. It can be constructed from an integral along the classical trajectory as
\be
T(b) = \exp \Bigg\{- \frac{2}{\hbar}\int_{-\infty}^{+\infty} U_I(R(t^\prime)) \> dt^\prime \Bigg\}
\ee
where $U_I$ is the imaginary part of the optical potential that describes the elastic scattering. The dependence on the impact parameter is implicitly contained in the relative trajectory $R(t)$. When the imaginary part is not available from the experimental data it is usual practice to set it as half the magnitude of the real part. In this semi-classical model, the optical potential determines the trajectory as well as part of the absorption. The cross section of Eq. (\ref{xsec}) is implicitly integrated over the full solid angle. However, most of the experimental data are taken within a finite-angle range. To make a quantitative comparison with the experimental data the integral in Eq. (\ref{xsec}) has to be reduced to the range of the impact parameters whose correlated trajectories  correspond to the experimental scattering-angle range. This can be achieved using the classical deflection function which relates the scattering angle with the impact parameter \cite{Lan14}.

Suppose the potential is weak and the Hamiltonian can be written as 
\be
H=H_{0} + \lambda W(t)
\ee
the perturbation is tuned by the dimensionless parameter $\lambda$, which can take the value from 0 (no perturbation) to 1 (maximum perturbation). Expanding the amplitudes $a_{\alpha}$ in powers of $\lambda$ and equating the terms with equal power, the first order gives
\be
a_{\alpha}^{(1)} (t) = - \frac{i}{\hbar} \int_{-\infty}^{t} < \phi_{\alpha} |W(t)| \phi_{\alpha^{\prime}}>
e^{\frac{i}{\hbar}\> (E_{\alpha} - E_{\alpha^{\prime}}) t^{\prime}} d\>t^{\prime}
\ee
with the initial condition $a(-\infty) = \delta_{\alpha, gs}$ and the integral is evaluated along the classical trajectory $R(t)$.

\section{\textit{Radial form factor}\label{ff}}

Whatever the method to solve the scattering problem is, the principal quantity that enters in the equations is the matrix element, which, for DWBA, was written as $< \psi_b \  \psi_B |  V |\ \psi_a \ \psi_A >$ for a reaction $A(a,b)B$ and whose integral form defining the form factor is
\be
F(r) = \int \psi_b \  \psi_B  \ V \ \psi_a \ \psi_A \ d\> x \>.
\ee
After the integration over all the internal coordinates of the collision partners and taking into account all the angular-momentum couplings, the form factor turns out to be a function only of the distance between the centres of mass of the two colliding ions and therefore is called radial form factor. This is the quantity that describes - sometimes in a very detailed way - the excitation process and it can be calculated in various ways depending on the model chosen. 

One of the most efficient ways to calculate it is by using the double-folding procedure, which has been successfully employed in the calculation of the ion-ion potential  \cite{Sat79,Sat83}. The method consists in choosing an effective nucleon-nucleon potential and integrating it over the ground-state densities of the two partners of the reaction. Taking into account also the isospin-dependent part, the effective  nucleon-nucleon interaction can be written as \cite{Sat83}
\be
v_{12} = v_0(r_{12}) + v_1(r_{12}) {\bf \tau_1 \cdot \tau_2} 
\ee 
where $v_{0}$ is the isoscalar term, $v_{1}$ the isovector part, and $\tau_{i}$ are the isospins of the nucleons.  With this choice, the neutron-neutron, proton-proton and neutron-proton interaction will be
\be
v_{nn}=v_{pp}=v_0+v_1 , \> v_{np}=v_0-v_1 .  
\ee 
Denoting with $\rho$ the ground-state densities of the two nuclei, then the definition of the double-folding potential gives
\bea \nonumber
U_F &=& \int\int \rho_a(r_1) \rho_A(r_2) v_{12} d{\bf r_1} d{\bf r_2} \\ \nonumber
 &=& \int\int \bigg[ \rho_{a_n}(r_1)\rho_{A_n}(r_2) +
\rho_{a_p}(r_1)\rho_{A_p}(r_2) \bigg] (v_0+v_1) d{\bf r_1} d{\bf r_2} \\ \nonumber
& \\
&+&\int\int \bigg[\rho_{a_n}(r_1)\rho_{A_p}(r_2) + \rho_{a_p}(r_1)\rho_{A_n}(r_2)\bigg]
 (v_0-v_1) d{\bf r_1} d{\bf r_2}
\eea 
where the contributions from neutrons ($n$) and protons ($p$) have been explicitly separated. The isoscalar and isovector parts of the interaction are 
\be
U_{F_0}=\int\int \rho_a(r_1) \rho_A(r_2) v_0(r_{12}) d{\bf r_1} d{\bf r_2} 
\ee 
\be
U_{F_1}=\int\int \{\rho_{a_n}(r_1)-\rho_{a_p}(r_1)\} 
\{\rho_{A_n}(r_2)-\rho_{A_p}(r_2)\} v_1(r_{12}) d{\bf r_1} d{\bf r_2}
\ee 
Note that in the particular case when $\rho_n=\rho_p (N/Z)= \rho (N/A) $ the last expression reduces to 
\be
U_{F_1}=\bigg( {{N_A-Z_A}\over A_A}\bigg) \bigg({{N_a-Z_a}\over A_a}\bigg)
\int\int \rho_{a}(r_1) \rho_{A}(r_1) v_1(r_{12}) d{\bf r_1} d{\bf
  r_2}.  
\ee
This implies that when one of the reaction partners has $N=Z$  the isovector part of the interaction $U_{F_1}$ is zero. Following the choice of Ref. \cite{Sat79} - the so-called M3Y nucleon-nucleon interaction, Reid type \cite{Ber77} - the explicit expressions for $v_0$ and $v_1$ (in MeV) and $r$ (in fm) are 
\be
v_0 (r)=\bigg[ 7999 {e^{-4r}\over 4r} - 2134 {e^{-2.5r}\over 2.5r} \bigg] 
- 262  \delta (\bf r)
\ee
and 
\be
v_1 (r)=-\bigg[ 4886 {e^{-4r}\over 4r} - 1176 {e^{-2.5r}\over 2.5r} \bigg] 
+ 217  \delta (\bf r) \> ,
\ee
Here, the zero-range term is the pseudo-potential, which takes into account, in an effective way, the single nucleon exchange \cite{Sat79}. 

The same procedure can be followed to build up the form factors where one uses the ground-state density of one nucleus and the transition density corresponding to the excited state of the other nucleus. Their formal expression can be obtained by proceeding as for the ion-ion potential
\be
F_{0}=\int\int \Big[\delta\rho_{a_n}(r_1) +  \delta\rho_{a_p}(r_1)\Big]
v_0(r_{12}) \Big[\rho_{A_p}(r_2)+\rho_{A_n}(r_2)\Big] r_1^2 \,dr_1 \,r_2^2 \,dr_2 
\label{f0} 
\ee
\be
F_{1}=\int\int \Big[\delta\rho_{a_n}(r_1) -  \delta\rho_{a_p}(r_1)\Big]
v_1(r_{12}) \Big[\rho_{A_n}(r_2)-\rho_{A_p}(r_2)\Big] r_1^2 \,dr_1 \,r_2^2 \,dr_2  
\label{f1} 
\ee
note that, as mentioned above, if one of the two nuclei has $N=Z$ then the isovector part $F_1$ goes to zero.
Any of the transition densities discussed in the previous section can be employed for the calculation of the form factors.
\begin{figure}[!htb]
\begin{center}
\includegraphics[width=8cm, angle=0]{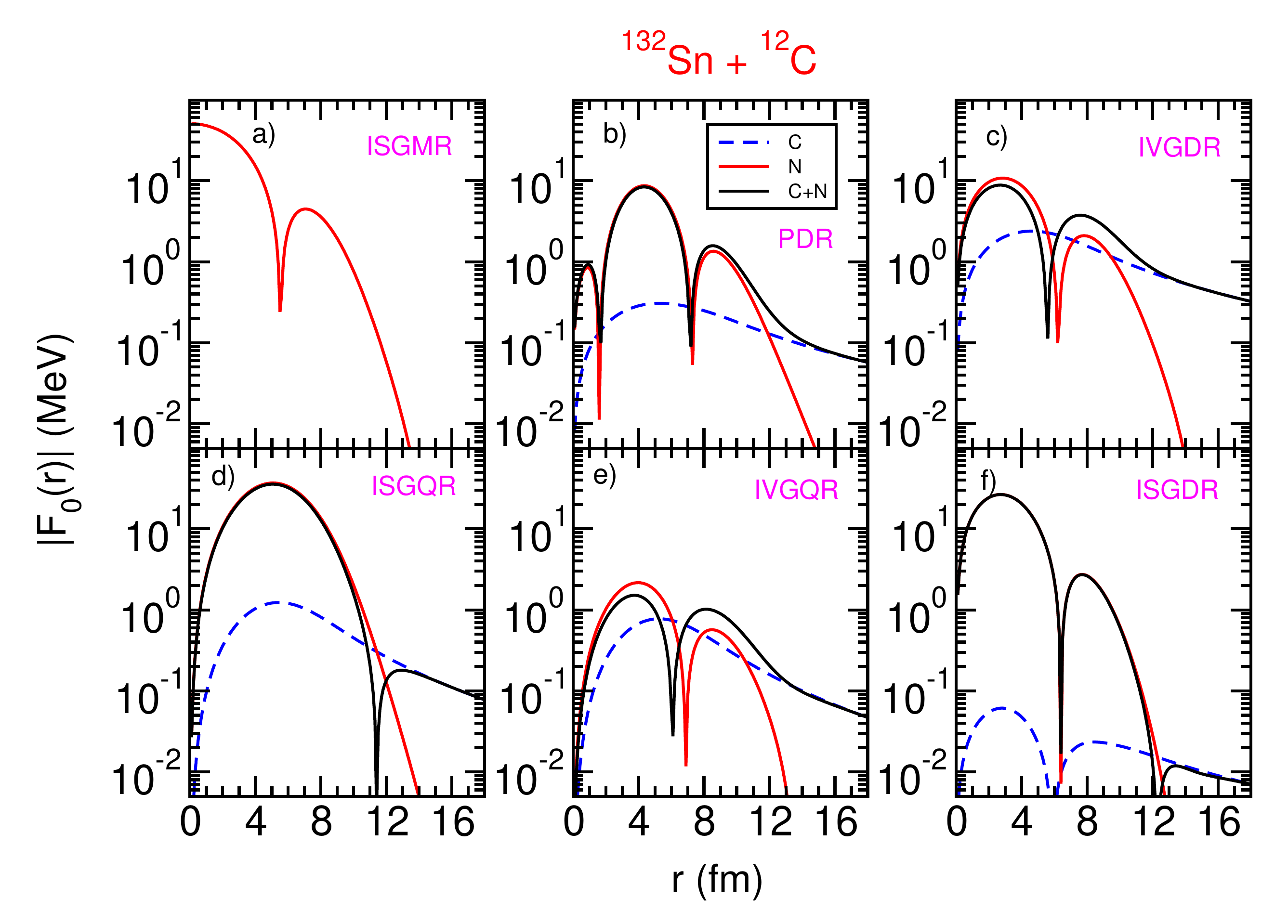}
\caption{(Color online)  Absolute values of the radial form factors for the system $^{132}$Sn + $^{12}$C. Each panel refers to an excited state of $^{132}$Sn. The Coulomb (dashed blue line) and nuclear (solid red line) are plotted separately. The black solid line is the result when the two contributions are taken into account simultaneously.}
\label{ff-132sn12c}
\end{center}
\end{figure}

Already from the radial shape of the form factors it is possible to infer some important information on the excitation process of a state. In fact, the form factor enters into the determination of the excitation cross section as its squared modulus. The isoscalar or isovector nature of the state determines the relative importance of the Coulomb and nuclear interactions in the excitation process. To illustrate this point, the radial form factors for several states of $^{132}$Sn excited by a $^{12}$C target are shown in Fig. \ref{ff-132sn12c}. The states were obtained performing a HF + RPA calculation \cite{Lan09} with a SGII \cite{Gia81,Gia81a} effective Skyrme interaction. The form factors were calculated with the double-folding procedure described above with the microscopic transition densities derived within the RPA. The isovector component $F_{1}$ is zero because the target $^{12}$C has $N=Z$. The Isoscalar Giant Monopole Resonance (ISGMR) is excited only by the nuclear potential as expected. The Isoscalar and Isovector Giant Quadrupole Resonances (ISGQR, IVGQR) have almost the same Coulomb contribution but a very different nuclear one. The PDR and IVGDR show the same nuclear contribution - in the grazing interaction region for these kinds of reactions - while the Coulomb contribution is very different: they have the same shape but almost an order of magnitude less for the PDR. The contribution of the Coulomb interaction to the ISGDR, which is a pure isoscalar dipole mode, is basically zero. The change in sign in the form factors depend on the nodes present in the transition densities. 
\begin{figure}[!htb]
\begin{center}
\includegraphics[width=11cm, angle=0]{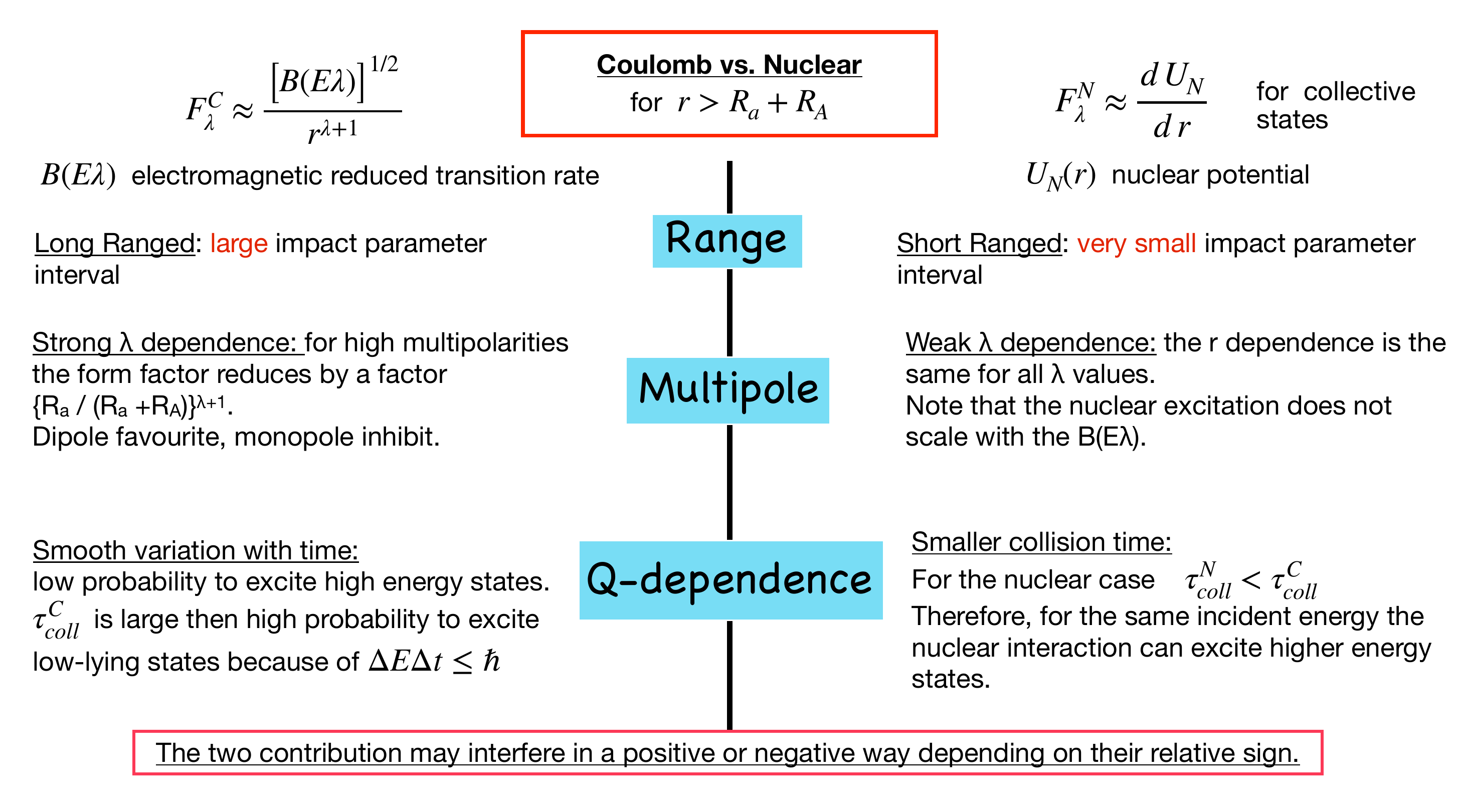}
\caption{(Color online)  Coulomb vs. nuclear properties.}
\label{ff-cou-nuc}
\end{center}
\end{figure}

Within the vibrational collective model it is possible to derive analytical expressions for the Coulomb and nuclear form factors \cite{Lan84}
\be
F_{\lambda} (r) = F_{\lambda}^{C} (r) + F_{\lambda}^{N} (r) = \frac{4 \pi\> Z_{a} \>e}{2 \lambda +1} 
\frac{[B(E \lambda)]^{1/2}}{r^{\lambda + 1}} - \delta_{\lambda}^{N} \>  \frac{d U^{N}}{dr}
\label{ff-macro}
\ee
where $\delta_{\lambda}^{N}$ is the nuclear deformation length. For many multipole states, the macroscopic and microscopic nuclear form factors are similar in the very peripheral nuclear region \cite{Lan84}. Actually, the Coulomb ones are identical for distances $r > R_{a}+R_{A}$. Therefore, some general properties of the form factors can be deduced from the expression (\ref{ff-macro}) and they are summarised in Fig. \ref{ff-cou-nuc}. Due to their $r$ dependence, the range of impact parameters giving substantial contributions to the Coulomb and nuclear form factors is very different. The Coulomb strongly depends on the multipolarity of the excited state while the nuclear one depends very weakly on it through the deformation length without changing its radial form. Most important probably is the Q-dependence, which prevents high-energy states to be excited by the Coulomb interaction while it has a high probability to excite low-lying states due to its large collision time. On the contrary, the nuclear collision time is very short and therefore it may excite higher energy states. The Coulomb and nuclear contributions to the inelastic cross section may interfere constructively or destructively depending on their relative sign.

\begin{figure}[!htb]
\begin{center}
\includegraphics[width=10cm, angle=0]{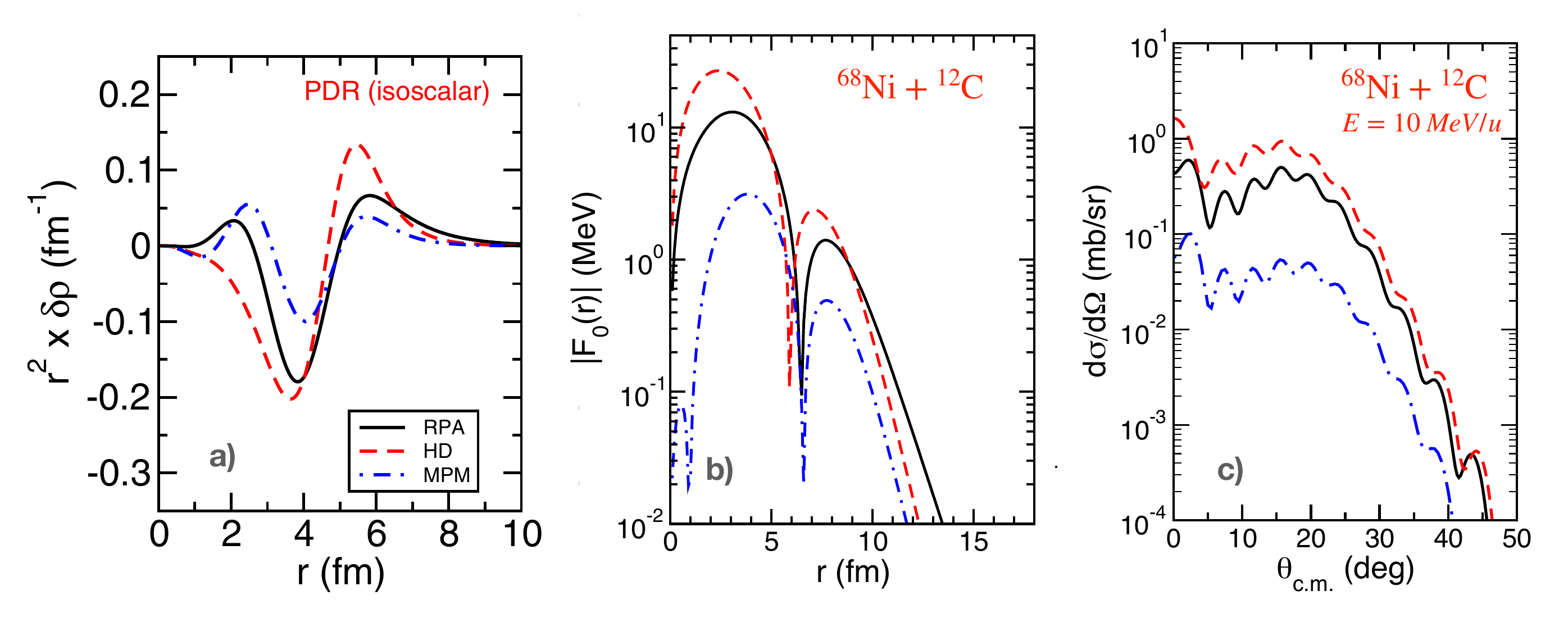}
\caption{ (Color online) In panel a), the isoscalar RPA transition density for $^{68}$Ni is compared with the one calculated for the ISGDR in Ref. \cite{Har81} (HD) and the one of the Macroscopic Pygmy Model (MPM) of Eq. (\ref{tdour-is}). In panel b), the form factors calculated for the system $^{68}$Ni + $^{12}$C with the double-folding procedure and with the transition densities of panel a). In panel c), the DWBA differential cross sections for the reaction $^{68}$Ni + $^{12}$C at 10 MeV/u, calculated with the form factors of panel b).
\label{td-ff-xsec}}
\end{center}
\end{figure}
A comparison among the different approaches to calculate the form factors should be performed in order to establish the appropriate one to be used in the cross section calculations. The three different form factors described above - microscopic RPA, the pure isoscalar of Ref. \cite{Har81} (HD) and the Macroscopic Pygmy Model (MPM) of Eq. (\ref{tdour-is}) \cite{Lan15} - are compared in panel b) of Fig. \ref{td-ff-xsec} for the system $^{68}$Ni + $^{12}$C. They are calculated with the double-folding procedure using the isoscalar transition densities plotted in panel a). The DWBA differential cross sections for the reaction $^{68}$Ni + $^{12}$C at 10 MeV/u in panel c) - calculated with the form factors of panel b) - are very different especially in the more important forward angular region. The factor-of-three difference between the RPA result and the HD one became almost an order of magnitude when the comparison is made with the macroscopic approach MPM. The DWBA calculations were performed with the DWUCK4 code \cite{Kunz, Kun93}. The RPA microscopic form factors have succeeded in the analysis of several experiments performed with $\alpha$ and $^{17}$O as isoscalar probe (see \cite{Bra15,Bra19} and references therein). 

\section{\textit{Coulomb and nuclear interplay}\label{cou-nuc}} 

\begin{figure}[!htb]
\begin{center}
\includegraphics[width=6cm, angle=0]{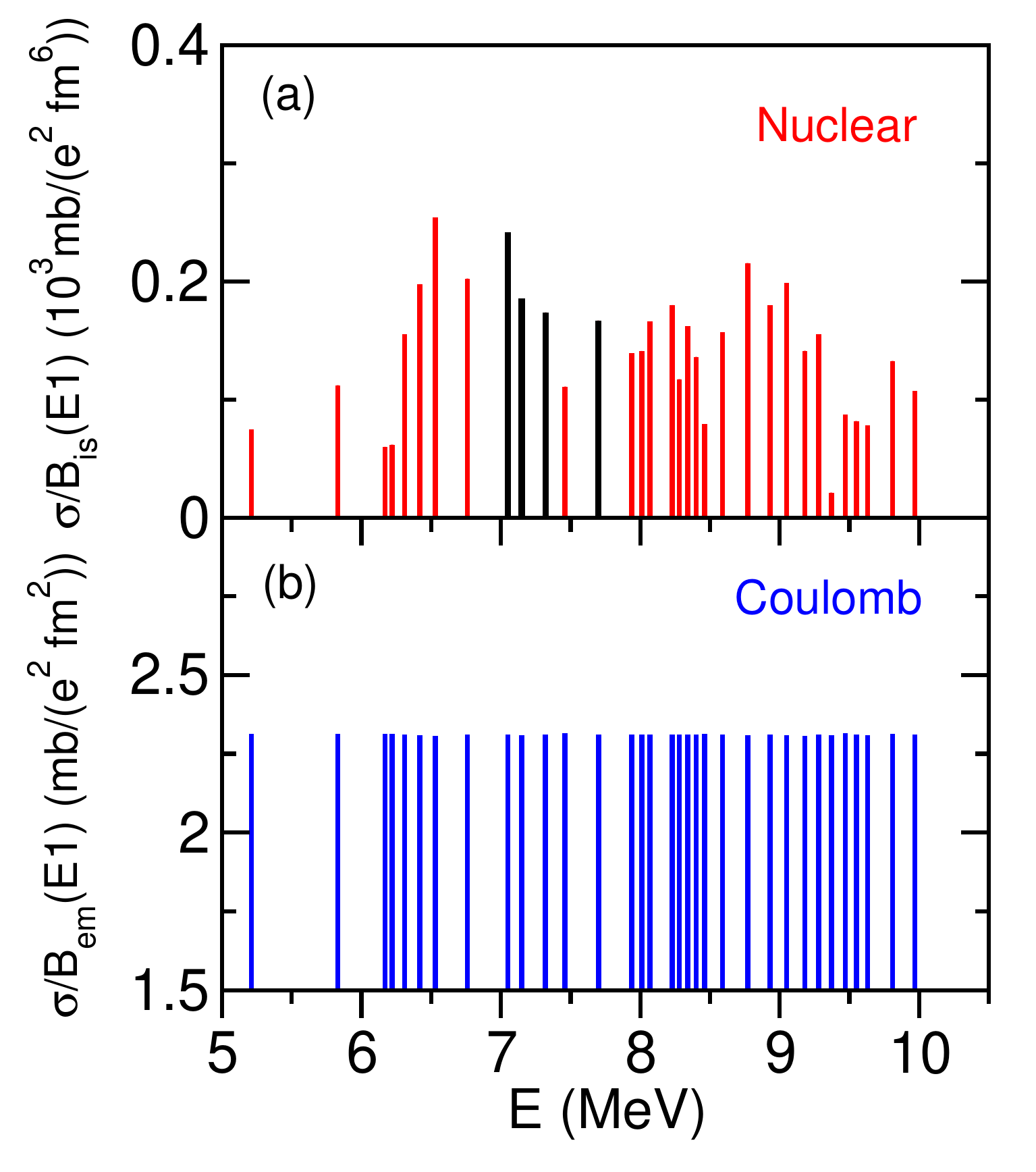}
\caption{(Color online) Ratios between the inelastic cross sections for the system $\alpha$ + $^{124}$Sn at 136 MeV incident energy, calculated by setting to zero the energies of the dipole states, and their corresponding B(EL). The two panels show the results for the nuclear (a) and the Coulomb (b) excitations. See text for more details. Taken with permission from \cite{Lan14}. \copyright 2014 by APS. \label{ratioxsec}}
\end{center}
\end{figure}

The different roles played by the Coulomb and nuclear interactions are manifested in the study of the isospin splitting of the PDR strength distribution \cite{End10} (see fig. \ref{fig-a-end10}). The relationship between the ($\alpha,\alpha^{\prime} \gamma$) inelastic cross section and the isoscalar $B_{is}(E1)$ is not so evident because - in contrast to the case of the Coulomb excitation cross section and the $B_{em}(E1)$ - there is no direct proportionality between the two quantities. This can be seen by calculating the Coulomb and nuclear inelastic cross sections separately and dividing them by the corresponding isoscalar and isovector B(E1) values. In Ref. \cite{Lan14}, this was done for the reaction $\alpha$ + $^{124}$Sn at 136 MeV incident energy \cite{End10}. However, the cross sections have extra factors related to the dynamics of the excitation process such as the energies of the states. Therefore, to avoid the biases due to this Q-value effect the cross section was calculated by imposing the energy of the dipole states used to be equal to zero.
In Fig.~\ref{ratioxsec}, the ratio
\begin{equation}
    {\sigma^i_\tau (E_i=0) \over B^i_\tau (E1)}
\end{equation}
is plotted for the nuclear (panel a) and Coulomb (panel b) cases. The $\sigma^i_\tau (E_i=0)$ is the excitation cross section for the dipole state $i$ calculated with its energy $E_i$ equal to zero and $B^i_\tau (E1)$ being its reduced transition probability. The index $\tau$ indicates the isoscalar or electromagnetic cases. The cross sections are calculated within the semi-classical model described above. These ratios are plotted at the values of the actual energies of the states. For the Coulomb case, the ratio is found to be constant for all the considered states as expected. For the nuclear excitation, the individual cross sections depend on the characteristics of the transition densities. Therefore, for nuclear excitations, the inelastic cross-section calculations have to be performed to have a precise comparison between experimental data and theoretical calculations. However, the calculated cross section reproduces the global features of the strong reduction in the experimental cross section at higher excitation energy compared to the isovector channel \cite{Lan14}.

\begin{figure}[!htb]
\begin{center}
\includegraphics[width=10cm, angle=0]{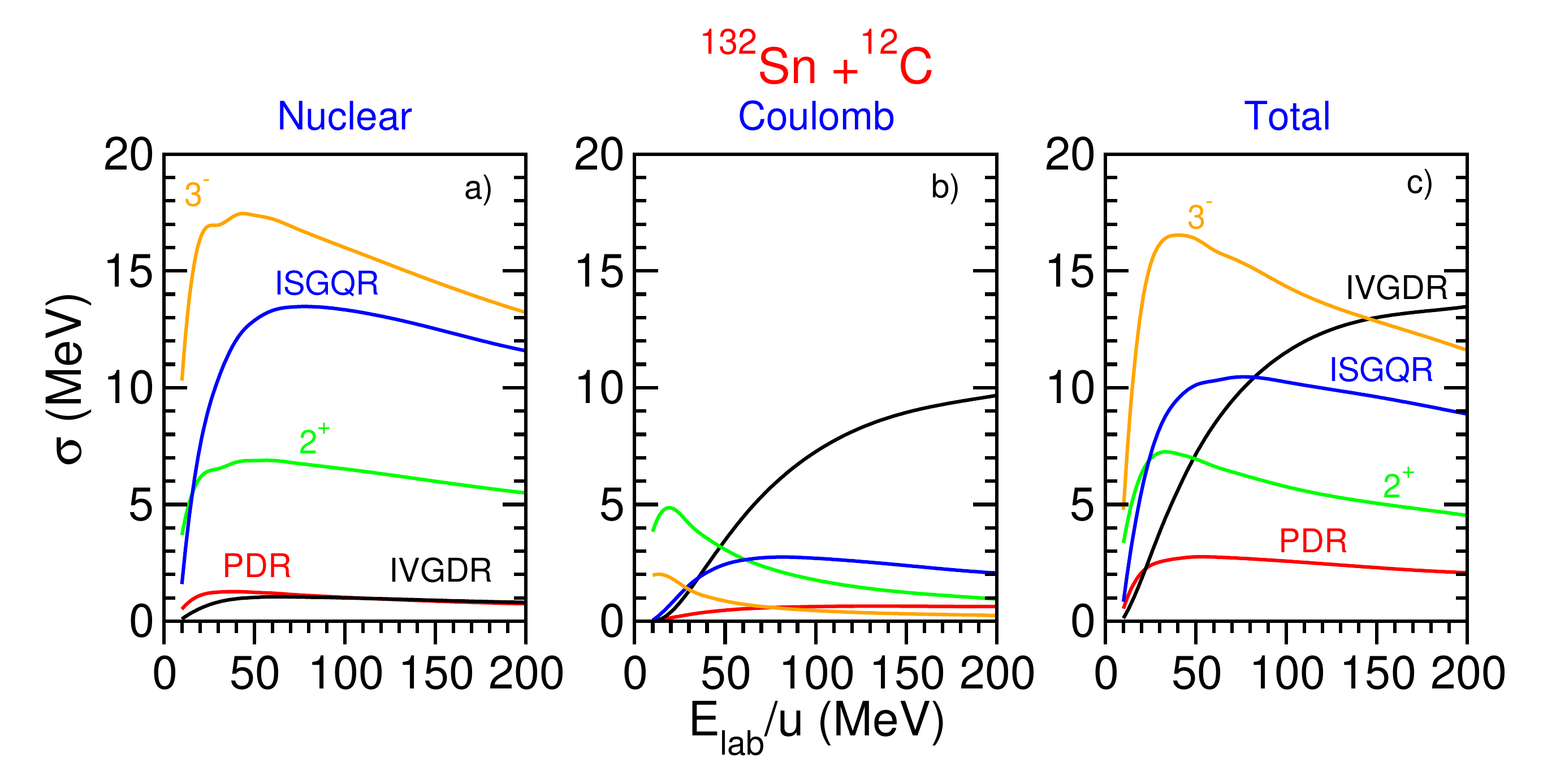}
\caption{(Color online) Excitation cross sections for the system $^{132}$Sn + $^{12}$C as a function of the incident energy per nucleon for the multipole states indicated in the figure. The cross sections due to the nuclear (a) and Coulomb (b) interactions, as well as the total one (c), are shown in separate frames. Each line corresponds to the cross section of the multipole state indicated by the label. \label{xsec-132sn12c}}
\end{center}
\end{figure}

The relative role of the nuclear and Coulomb components in the inelastic cross section can be modified by appropriately tuning the projectile mass, charge, bombarding energy and scattering angle together with the isoscalar and isovector contributions. An example of how the inelastic cross section can change with the incident energy and for states of different multipoles is shown in Fig. \ref{xsec-132sn12c} where the excitation cross section for the system $^{132}$Sn + $^{12}$C is plotted as function of the incident energy per nucleon. The excited states of $^{132}$Sn taken in considerations are calculated with a HF + RPA with a SGII Skyrme interaction following the procedure of Ref. \cite{Lan09}. The effect of the multipole dependence of the Coulomb form factors for the low-lying $2^{+}$ and $3^{-}$ states is clearly evident in panel b) as well as the Q-dependence effect. The excitation due to the nuclear interaction favours isoscalar states (panel a) as well as the PDR due to its isospin mixing, especially at low incident energies. Taking into consideration both interactions a strong positive interference is achieved for the dipole states. According to these results, the study of the PDR with isoscalar probes is more effective at incident energies around 30 MeV per nucleon. Most of the experiment using an isoscalar probe like $\alpha$ particle, $^{17}$O or $^{12}$C have been performed at an incident energy within the range suggested by the calculations shown above, with very satisfying results (see the review papers \cite{Sav13,Bra15,Bra19} and references therein for more details).

\section{\textit{Summary}\label{summ}}

Most of the theoretical works investigating the Pygmy Dipole Resonances are dedicated to the interpretation of the observed characteristics of the low-lying dipole strength distribution. Macroscopic models based on the generalised Goldhaber-Teller or Steinwedel-Jensen models describe this excitation as due to a three incompressible fluids: the excess neutrons, forming a {\it neutron skin}, oscillate against an isospin inert core giving rise to a dipole excitation whose energy is found to be around the neutron emission threshold. This interpretation gave rise to the classical, schematic picture of the PDR that is commonly used. The physical insight provided by the macroscopic models is improved by the use of sophisticated microscopic models like RPA and its extensions. In the first part of this chapter, after a brief report on the main experimental findings, a description of the theoretical models employed in the studies of the PDR is given. Only the main aspects and results regarding the PDR are underlined, referring to more detailed descriptions of these approaches in text books \cite{Row10,Rin04}. It may be also fruitful to read the following reviews \cite{Paa07,Roc18,Lan22}.

The RPA approach describes the excitation of closed-shell nuclei and already at this level the main features of the PDR are well predicted. The use of non-relativistic and relativistic QRPA allows the description of open-shell and deformed nuclei. Extensions of the RPA by including the coupling of the elementary $1p-1h$ configuration with more complex ones, as second RPA (SRPA) or subtracted SRPA (SSRPA), allow a more detailed description of the low-lying dipole excitations, such as the fragmentation in numerous states with a centroid positioned at a lower energy. Finally, the coupling with two- and three-phonon states, achieved in the QPM and RQTBA approaches, gives a good description of the PDR below the neutron emission threshold by reproducing the strength fragmentation and the isospin splitting. All these theoretical approaches, from the more simple to the more sophisticated ones, agree in defining the main features of the Pygmy Dipole Resonances. Combining these results with the experimental findings, the main characteristics of the PDR can be summarised as follow. They are found in neutron-rich stable and unstable nuclei exhausting a few percent of EWSR and at energies around the neutron emission threshold. They have a strong isospin mixing allowing the investigation with both isoscalar and isovector probes. All the theoretical approaches show in fact that the transition densities of the PDR states present a strong isospin mixing at the nuclear surface. The isospin splitting of the PDR - identified when isoscalar probes are also used in their studies - can be reproduced by theoretical calculations when the coupling to two- and three-phonon states are taken into account. It became evident in the past years that more detailed information on the nature of this excitation mode can be extracted from the use of complementary probes rather than limiting the study to electromagnetic probes. It is therefore important to have also theoretical models to calculate the corresponding inelastic cross section. Most of the theoretical studies of the PDR are dedicated to describe and understand the structure of the dipole strength distribution while very few contributions are focused on the inelastic cross-section calculations. To fill partially this gap, the second part of this chapter is dedicated to illustrate a semi-classical coupled-channel model that has been used in many other nuclear physics fields. To get reliable calculations of the inelastic cross section, it is of paramount importance to have form factors that describe in a realistic way the excitation process. The double-folding procedure using microscopic transition densities seems to be the most reliable one as proved by their use in obtaining a good description of many experimental data. The interplay between Coulomb and nuclear excitation is crucial in the comprehension of the PDR mode and cross-section calculations show that the study of the PDR with  isoscalar probes is best achieved at low incident energies (around 30 MeV/u).

Nowadays, all the different theoretical models are able to describe and reproduce most of the experimental observables obtained from the experiments performed to study the PDR, starting from the strength distribution to the inelastic excitation process. Despite of the substantial progress achieved in this direction, there are still some aspects that need to be clarified. The debate whether the PDR mode is collective or not is still far from resolved. Some future experiments are planned to provide answers to this problem. The interplay between the isoscalar and isovector responses below and above the neutron emission threshold has to be investigated to get new information especially for radioactive nuclei far from the stability valley. More attention should also be given to the role that nuclear deformation plays in the excitation of the low-lying dipole mode especially addressing the contradictory results found by different theoretical calculations.

\bibliography{mybibfile}

\begin{thebibliography}{}

\bibitem[Adrich et~al., 2005]{Adr05}
Adrich, P. et~al. (2005).
\newblock {\em Phys. Rev. Lett.}, 95:132501.

\bibitem[Allison et~al., 2016]{GEANT4}
Allison, J. et~al. (2016).
\newblock {\em Nuclear Instruments and Methods in Physics Research Section A:
  Accelerators, Spectrometers, Detectors and Associated Equipment},
  835:186--225.

\bibitem[Aumann, 2019]{Aum19}
Aumann, T. (2019).
\newblock {\em The European Physical Journal A}, 55:234.

\bibitem[Berger et~al., 1991]{Ber91}
Berger, J., Girod, M., and Gogny, D. (1991).
\newblock {\em Comp. Phys. Comm.}, 63:365.

\bibitem[Bertsch et~al., 1977]{Ber77}
Bertsch, G., Borysowicz, J., McManus, H., and Love, W. (1977).
\newblock {\em Nuclear Physics A}, 284:399--419.

\bibitem[Bertulani and Ponomarev, 1999]{Ber99}
Bertulani, C. and Ponomarev, V. (1999).
\newblock {\em Physics Reports}, 321:139.

\bibitem[Bohr and Mottelson, 1975]{Boh75}
Bohr, A. and Mottelson, B. (1975).
\newblock {\em Nuclear Structure,}.
\newblock W.A. Benjamin, New York,.

\bibitem[Bortignon et~al., 2019]{Bor19}
Bortignon, P.~F., Bracco, A., and Broglia, R.~A. (2019).
\newblock {\em Giant Resonances: Nuclear structure at finite temperature}.
\newblock CRC Press.

\bibitem[Bracco et~al., 2015]{Bra15}
Bracco, A., Crespi, F., and Lanza, E. (2015).
\newblock {\em Eur Phys. J. A 51}, page~99.

\bibitem[Bracco et~al., 2019]{Bra19}
Bracco, A., Lanza, E.~G., and Tamii, A. (2019).
\newblock {\em Prog. Part. Nucl. Phys.}, 106:360.

\bibitem[Broglia and Winther, 2004]{Bro04}
Broglia, R. and Winther, A. (2004).
\newblock {\em Heavy Ion Reactions: The Elementary Processes, Part I and II,}.
\newblock CRC Press.

\bibitem[Broglia and Winther, 1972]{Bro72}
Broglia, R.~A. and Winther, A. (1972).
\newblock {\em Phys. Rep.}, 4:153.

\bibitem[Catara et~al., 1997a]{Cat97a}
Catara, F., Lanza, E.~G., Nagarajan, M.~A., and Vitturi, A. (1997a).
\newblock {\em Nucl. Phys. A}, 614:86.

\bibitem[Catara et~al., 1997b]{Cat97b}
Catara, F., Lanza, E.~G., Nagarajan, M.~A., and Vitturi, A. (1997b).
\newblock {\em Nucl. Phys. A}, 624:449.

\bibitem[Decharg\'e and Gogny, 1980]{Dec80}
Decharg\'e, J. and Gogny, D. (1980).
\newblock {\em Phys. Rev. C}, 21:1568--1593.

\bibitem[Enders et~al., 2000]{End00}
Enders, J., {von Brentano}, P., Eberth, J., Fitzler, A., Fransen, C., Herzberg,
  R.-D., Kaiser, H., Käubler, L., {von Neumann-Cosel}, P., Pietralla, N.,
  Ponomarev, V., Prade, H., Richter, A., Schnare, H., Schwengner, R., Skoda,
  S., Thomas, H., Tiesler, H., Weisshaar, D., and Wiedenhöver, I. (2000).
\newblock {\em Physics Letters B}, 486:279--285.

\bibitem[Enders et~al., 2003]{End03}
Enders, J., von Brentano, P., Eberth, J., Fitzler, A., Fransen, C., Herzberg,
  R.-D., Kaiser, H., Käubler, L., von Neumann Cosel, P., Pietralla, N.,
  Ponomarev, V., Richter, A., Schwengner, R., and Wiedenhöver, I. (2003).
\newblock {\em Nuclear Physics A}, 724:243--273.

\bibitem[Endres et~al., 2010]{End10}
Endres, J. et~al. (2010).
\newblock {\em Phys. Rev. Lett.}, 105:212503.

\bibitem[Fallot et~al., 2003]{Fal03}
Fallot, M., Chomaz, P., Andrés, M., Catara, F., Lanza, E., and Scarpaci, J.
  (2003).
\newblock {\em Nuclear Physics A}, 729:699--712.

\bibitem[Gambacurta et~al., 2011]{Gam11}
Gambacurta, D., Grasso, M., and Catara, F. (2011).
\newblock {\em Phys. Rev. C}, 84:034301.

\bibitem[Gambacurta et~al., 2018]{Gam18}
Gambacurta, D., Grasso, M., and Vasseur, O. (2018).
\newblock {\em Physics Letters B}, 777:163--168.

\bibitem[Giai and Sagawa, 1981a]{Gia81}
Giai, N.~V. and Sagawa, H. (1981a).
\newblock {\em Phys. Lett. B}, 106:379.

\bibitem[Giai and Sagawa, 1981b]{Gia81a}
Giai, N.~V. and Sagawa, H. (1981b).
\newblock {\em Nucl. Phys. A}, 371:1.

\bibitem[Goddard et~al., 2013]{God13}
Goddard, P.~M., Cooper, N., Werner, V., Rusev, G., Stevenson, P.~D., Rios, A.,
  Bernards, C., Chakraborty, A., Crider, B.~P., Glorius, J., Ilieva, R.~S.,
  Kelley, J.~H., Kwan, E., Peters, E.~E., Pietralla, N., Raut, R., Romig, C.,
  Savran, D., Schnorrenberger, L., Smith, M.~K., Sonnabend, K., Tonchev, A.~P.,
  Tornow, W., and Yates, S.~W. (2013).
\newblock {\em Phys. Rev. C}, 88:064308.

\bibitem[Goldhaber and Teller, 1948]{Gol48}
Goldhaber, M. and Teller, E. (1948).
\newblock {\em Phys. Rev.}, 74:1046.

\bibitem[Grasso and Gambacurta, 2020]{Gam20}
Grasso, M. and Gambacurta, D. (2020).
\newblock {\em Phys. Rev. C}, 101:064314.

\bibitem[Hamamoto and Sagawa, 1996]{Ham96a}
Hamamoto, I. and Sagawa, H. (1996).
\newblock {\em Phys. Rev. C}, 53:R1492.

\bibitem[Hamamoto et~al., 1996]{Ham96}
Hamamoto, I., Sagawa, H., and Zhang, X.~Z. (1996).
\newblock {\em Phys. Rev. C}, 53:765.

\bibitem[Hamamoto et~al., 1998]{Ham98}
Hamamoto, I., Sagawa, H., and Zhang, X.~Z. (1998).
\newblock {\em Phys. Rev. C}, 57:R1064.

\bibitem[Harakeh and Dieperink, 1981]{Har81}
Harakeh, M.~N. and Dieperink, A. E.~L. (1981).
\newblock {\em Phys. Rev. C}, 23:2329--2334.

\bibitem[Harakeh and van~der Woude, 2001]{Har01}
Harakeh, M.~N. and van~der Woude, A. (2001).
\newblock {\em Giant Resonances,}.
\newblock Clarendon Press, Oxford.

\bibitem[Iachello and Arima, 1987]{Iac87}
Iachello, F. and Arima, A. (1987).
\newblock {\em The Interacting Boson Model,}.
\newblock Cambridge University Press,.

\bibitem[Isacker et~al., 1992]{Van92}
Isacker, P.~V., Nagarajan, M., and Warner, D. (1992).
\newblock {\em Phys. Rev. C}, 45:R13.

\bibitem[Klimkiewicz et~al., 2007]{Kli07}
Klimkiewicz, A. et~al. (2007).
\newblock {\em Nucl. Phys. A}, 788:145.

\bibitem[Kunz, ]{Kunz}
Kunz, P.
\newblock {\em DWUCK4 code for DWBA}, at
  https://people.nscl.msu.edu/~brown/reaction-codes/.

\bibitem[Kunz and Rost, 1993]{Kun93}
Kunz, P. and Rost, E. (1993).
\newblock {\em The Distorted-Wave Born Approximation. In: Computational Nuclear
  Physics 2.}
\newblock Springer, New York, NY.

\bibitem[Kvasil et~al., 2011]{Kva11}
Kvasil, J., Nesterenko, V.~O., Kleinig, W., Reinhard, P.-G., and Vesely, P.
  (2011).
\newblock {\em Phys. Rev. C}, 84:034303.

\bibitem[Lalazissis et~al., 1997]{Lal97}
Lalazissis, G.~A., K\"onig, J., and Ring, P. (1997).
\newblock {\em Phys. Rev. C}, 55:540--543.

\bibitem[Landowne and Vitturi, 1984]{Lan84}
Landowne, S. and Vitturi, A. (1984).
\newblock {\em in Treatise on Heavy-Ion Science, edited by D. A. Bromley,}.
\newblock Plenum, New York,.

\bibitem[Lanza et~al., 1997]{Lan97}
Lanza, E., Andrés, M., Catara, F., Chomaz, P., and Volpe, C. (1997).
\newblock {\em Nuclear Physics A}, 613:445--471.

\bibitem[Lanza et~al., 2006]{Lan06}
Lanza, E.~G., Catara, F., Andr\'es, M.~V., Chomaz, P., Fallot, M., and
  Scarpaci, J.~A. (2006).
\newblock {\em Phys. Rev. C}, 74:064614.

\bibitem[Lanza et~al., 2009]{Lan09}
Lanza, E.~G., Catara, F., Gambacurta, D., Andr\'es, M.~V., and Chomaz, P.
  (2009).
\newblock {\em Phys. Rev. C}, 79:054615.

\bibitem[Lanza et~al., 2022]{Lan22}
Lanza, E.~G., Pellegri, L., Vitturi, A., and Andr\'es, M.~V. (2022).
\newblock {\em Prog. Part. Nucl. Phys.}

\bibitem[Lanza et~al., 2015]{Lan15}
Lanza, E.~G., Vitturi, A., and Andr\'es, M.~V. (2015).
\newblock {\em Phys. Rev. C}, 91:054607.

\bibitem[Lanza et~al., 2014]{Lan14}
Lanza, E.~G., Vitturi, A., Litvinova, E., and Savran, D. (2014).
\newblock {\em Phys. Rev. C}, 89:041601(R).

\bibitem[Liang et~al., 2007]{Lia07}
Liang, J., Cao, L.-G., and Ma, Z.-Y. (2007).
\newblock {\em Phys. Rev. C}, 75:054320.

\bibitem[Litvinova et~al., 2007]{Lit07}
Litvinova, E., Ring, P., and Tselyaev, V. (2007).
\newblock {\em Phys. Rev. C}, 75:064308.

\bibitem[Litvinova et~al., 2008]{Lit08}
Litvinova, E., Ring, P., and Tselyaev, V. (2008).
\newblock {\em Phys. Rev. C}, 78:014312.

\bibitem[Litvinova et~al., 2009]{Lit09}
Litvinova, E., Ring, P., Tselyaev, V., and Langanke, K. (2009).
\newblock {\em Phys. Rev. C}, 79:054312.

\bibitem[Martorana et~al., 2018]{Mar18}
Martorana, N. et~al. (2018).
\newblock {\em Phys. Lett. B}, 782:112.

\bibitem[Nesterenko et~al., 2016]{Nes16}
Nesterenko, V.~O., Kvasil, J., Repko, A., Kleinig, W., and Reinhard, P.~G.
  (2016).
\newblock {\em Physics of Atomic Nuclei}, 79:842--850.

\bibitem[Nik\^si\'c et~al., 2002]{Nik02}
Nik\^si\'c, T., Vretenar, D., and Ring, P. (2002).
\newblock {\em Phys. Rev. C}, 66:064302.

\bibitem[Paar et~al., 2003]{Paa03}
Paar, N., Ring, P., Nik\ifmmode \check{s}\else
  \v{s}\fi{}i\ifmmode~\acute{c}\else \'{c}\fi{}, T., and Vretenar, D. (2003).
\newblock {\em Phys. Rev. C}, 67:034312.

\bibitem[Paar et~al., 2007]{Paa07}
Paar, N., Vretenar, D., Khan, E., and Col{\`{o}}, G. (2007).
\newblock {\em Reports on Progress in Physics}, 70:691--793.

\bibitem[Papakonstantinou et~al., 2012]{Pap12}
Papakonstantinou, P., Hergert, H., Ponomarev, V.~Y., and Roth, R. (2012).
\newblock {\em Phys. Lett. B}, 709:270.

\bibitem[Papakonstantinou et~al., 2014]{Pap14}
Papakonstantinou, P., Hergert, H., Ponomarev, V.~Y., and Roth, R. (2014).
\newblock {\em Phys. Rev. C}, 89:034306.

\bibitem[Papakonstantinou and Roth, 2009]{Pap09}
Papakonstantinou, P. and Roth, R. (2009).
\newblock {\em Physics Letters B}, 671:356--360.

\bibitem[Pascu et~al., 2012]{Pas12}
Pascu, S., Endres, J., Zamfir, N.~V., and Zilges, A. (2012).
\newblock {\em Phys. Rev. C}, 85:064315.

\bibitem[Pe\~{n}a Arteaga et~al., 2009]{Pen09}
Pe\~{n}a Arteaga, D., Khan, E., and Ring, P. (2009).
\newblock {\em Phys. Rev. C}, 79:034311.

\bibitem[Pe\~{n}a Arteaga and Ring, 2008]{Pen08}
Pe\~{n}a Arteaga, D. and Ring, P. (2008).
\newblock {\em Phys. Rev. C}, 77:034317.

\bibitem[Piekarewicz, 2006]{Pie06}
Piekarewicz, J. (2006).
\newblock {\em Phys. Rev. C}, 73:044325.

\bibitem[Piekarewicz, 2011]{Pie11}
Piekarewicz, J. (2011).
\newblock {\em Phys. Rev. C}, 83:034319.

\bibitem[Repko et~al., 2019]{Rep19}
Repko, A., Nesterenko, V.~O., Kvasil, J., and Reinhard, P.~G. (2019).
\newblock {\em The European Physical Journal A}, 55:242.

\bibitem[Repko et~al., 2013]{Rep13}
Repko, A., Reinhard, P.-G., Nesterenko, V.~O., and Kvasil, J. (2013).
\newblock {\em Phys. Rev. C}, 87:024305.

\bibitem[Ring and Schuck, 2004]{Rin04}
Ring, P. and Schuck, P. (2004).
\newblock {\em The nuclear many-body problem}.
\newblock Springer,.

\bibitem[Roca-Maza and Paar, 2018]{Roc18}
Roca-Maza, X. and Paar, N. (2018).
\newblock {\em Prog. Part. Nucl. Phys.}, 101:96.

\bibitem[Rossi et~al., 2013]{Ros13}
Rossi, D. et~al. (2013).
\newblock {\em Phys. Rev. Lett.}, 111:242503.

\bibitem[Rowe, 2010]{Row10}
Rowe, D.~J. (2010).
\newblock {\em Nuclear collective motion,}.
\newblock World Scientific Publishing,.

\bibitem[Ryezayeva et~al., 2002]{Rye02}
Ryezayeva, N. et~al. (2002).
\newblock {\em Phys. Rev. Lett.}, 89:272502.

\bibitem[Satchler, 1983]{Sat83}
Satchler, G. (1983).
\newblock {\em Direct Nuclear Reactions}.
\newblock Oxford University Press,.

\bibitem[Satchler and Love, 1979]{Sat79}
Satchler, G. and Love, W. (1979).
\newblock {\em Physics Reports}, 55:183--254.

\bibitem[Savran et~al., 2013]{Sav13}
Savran, D., Aumann, T., and Zilges, A. (2013).
\newblock {\em Prog. Part. Nucl. Phys.}, 70:210.

\bibitem[Savran et~al., 2011]{Sav11}
Savran, D., Elvers, M., Endres, J., Fritzsche, M., L\"oher, B., Pietralla, N.,
  Ponomarev, V.~Y., Romig, C., Schnorrenberger, L., Sonnabend, K., and Zilges,
  A. (2011).
\newblock {\em Phys. Rev. C}, 84:024326.

\bibitem[Skyrme, 1958]{Sky58}
Skyrme, T. (1958).
\newblock {\em Nucl. Phys.}, 9:615.

\bibitem[Soloviev, 1992]{Sol92}
Soloviev, V. (1992).
\newblock {\em Theory of Atomic Nucleus: Quasiparticles and Phonons,}.
\newblock Institute of Physics, Bristol,.

\bibitem[Steinwedel and Jensen, 1950]{Ste50}
Steinwedel, H. and Jensen, J. H.~D. (1950).
\newblock {\em Z. Naturforsch.}, 5a:413.

\bibitem[Suzuki et~al., 1990]{Suz90}
Suzuki, Y., Ikeda, K., and Sato, H. (1990).
\newblock {\em Prog. Theor. Phys.}, 83:180.

\bibitem[Terasaki and Engel, 2006]{Ter06}
Terasaki, J. and Engel, J. (2006).
\newblock {\em Phys. Rev. C}, 74:044301.

\bibitem[Tselyaev, 2007]{Tse07}
Tselyaev, V.~I. (2007).
\newblock {\em Phys. Rev. C}, 75.

\bibitem[Tselyaev, 2013]{Tse13}
Tselyaev, V.~I. (2013).
\newblock {\em Phys. Rev. C}, 88:054301.

\bibitem[Tsoneva and Lenske, 2008]{Tso08}
Tsoneva, N. and Lenske, H. (2008).
\newblock {\em Phys. Rev. C}, 77:024321.

\bibitem[Tsoneva et~al., 2004]{Tso04}
Tsoneva, N., Lenske, H., and Stoyanov, C. (2004).
\newblock {\em Physics Letters B}, 586:213--218.

\bibitem[Urban, 2012]{Urb12}
Urban, M. (2012).
\newblock {\em Phys. Rev. C}, 85:034322.

\bibitem[Volz et~al., 2006]{Vol06}
Volz, S., Tsoneva, N., Babilon, M., Elvers, M., Hasper, J., Herzberg, R.-D.,
  Lenske, H., Lindenberg, K., Savran, D., and Zilges, A. (2006).
\newblock {\em Nuclear Physics A}, 779:1--20.

\bibitem[Vretenar et~al., 1995]{Vre95}
Vretenar, D., Berghammer, H., and Ring, P. (1995).
\newblock {\em Nucl. Phys. A}, 581:679.

\bibitem[Vretenar et~al., 2012]{Vre12}
Vretenar, D., Niu, Y.~F., Paar, N., and Meng, J. (2012).
\newblock {\em Phys. Rev. C}, 85:044317.

\bibitem[Vretenar et~al., 2001]{Vre01}
Vretenar, D., Paar, N., Ring, P., and Lalazissis, G. (2001).
\newblock {\em Nuclear Physics A}, 692:496--517.

\bibitem[Vretenar et~al., 2002]{Vre02}
Vretenar, D., Paar, N., Ring, P., and Nik\ifmmode \check{s}\else
  \v{s}\fi{}i\ifmmode~\acute{c}\else \'{c}\fi{}, T. (2002).
\newblock {\em Phys. Rev. C}, 65:021301.

\bibitem[Walecka, 1974]{Wal74}
Walecka, J.~D. (1974).
\newblock {\em Ann of Phys.}, 83:491.

\bibitem[Wieland et~al., 2009]{Wie09}
Wieland, O. et~al. (2009).
\newblock {\em Phys. Rev. Lett.}, 102:092502.

\bibitem[Yoshida and Nakatsukasa, 2011]{Yos11}
Yoshida, K. and Nakatsukasa, T. (2011).
\newblock {\em Phys. Rev. C}, 83:021304.

\bibitem[Zilges et~al., 2002]{Zil02}
Zilges, A., Volz, S., Babilon, M., Hartmann, T., Mohr, P., and Vogt, K. (2002).
\newblock {\em Physics Letters B}, 542:43--48.

\end{thebibliography}
\bibliographystyle{apalike}

\end{document}